\documentclass[aip,reprint,superscriptaddress,nofootinbib,twocolumn,10pt]{revtex4-2}

\usepackage[linktoc=page,colorlinks,urlcolor=blue,citecolor=blue,linkcolor=blue]{hyperref}
\usepackage{amssymb,amsmath}
\usepackage{upgreek}
\usepackage{graphicx}
\usepackage{natbib}
\usepackage{lineno}
\usepackage[dvipsnames]{xcolor}% \usepackage{mathrsfs}
\usepackage{overpic}
\usepackage{pdfpages}

\makeatletter
\AtBeginDocument{\let\LS@rot\@undefined}
\makeatother

\newcommand{\ket}[1]{|#1\rangle}
\newcommand{\snl}{Sandia National Laboratories, Albuquerque, New Mexico 87185, USA}
\newcommand{\cint}{Center for Integrated Nanotechnologies, Sandia National Laboratories, Albuquerque, New Mexico 87123, USA}
\newcommand{\llMIT}{Current address: MIT Lincoln Laboratory, Lexington, Massachusetts 02421, USA}

\usepackage{verbatim}

\begin{document}
\title{Wide-field microwave magnetic field imaging with nitrogen-vacancy centers in diamond}

\date{\today}
\author{Luca Basso}\email{lbasso@sandia.gov}\affiliation{\cint}
\author{Pauli Kehayias}\affiliation{\snl}\affiliation{\llMIT}
\author{Jacob Henshaw}\affiliation{\cint}\affiliation{\snl}
\author{Gajadhar Joshi}\affiliation{\cint}
\author{Michael P. Lilly}\affiliation{\cint}
\author{Matthew B. Jordan}\affiliation{\snl}
\author{Andrew M. Mounce}\email{ammounce@sandia.gov}\affiliation{\cint}
%TC:ignore

\begin{abstract}
Non-invasive imaging of microwave (MW) magnetic fields with microscale lateral resolution is pivotal for various applications, such as MW technologies and integrated circuit failure analysis. Diamond nitrogen-vacancy (NV) center magnetometry has emerged as an ideal tool, offering $\upmu$m-scale resolution, millimeter-scale field of view, high sensitivity, and non-invasive imaging compatible with diverse samples. However, up until now, it has been predominantly used for imaging of static or low-frequency magnetic fields or, concerning MW field imaging, to directly characterize the same microwave device used to drive the NV spin transitions. In this work we leverage an NV center ensemble in diamond for wide-field imaging of MW magnetic fields generated by a test device employing a differential measurement protocol. The microscope is equipped with a MW loop to induce Rabi oscillations between NV spin states, and the MW field from the device-under-test is measured through local deviations in the Rabi frequency. This differential protocol yields magnetic field maps of a 2.57 GHz MW field with a sensitivity of $\sim$ 9 $\upmu$T Hz$^{-1/2}$ for a total measurement duration of $T = 357$ s, covering a $340\times340$ $\upmu$m$^2$ field of view with a $\upmu$m-scale spatial resolution and a DUT input power dynamic range of 30 dB. This work demonstrates a novel NV magnetometry protocol, based on differential Rabi frequency measurement, that extends NV wide-field imaging capabilities to imaging of weak MW magnetic fields that would be difficult to measure directly through standard NV Rabi magnetometry.
\end{abstract}

\maketitle
\section{Introduction} 
The detection and imaging of microwave (MW) magnetic fields is of significant importance for a variety of applications, from probing magnetic excitations such as spin waves\cite{van2015nanometre} and spin liquids is frustrated magnets\cite{balents2010spin}, to mapping the AC susceptibility of magnetic materials\cite{dasika2023mapping}, as well as imaging of biological samples\cite{hall2013nanoscale}. Of particular interest is the use of MW imaging for MW devices characterization\cite{sayil2005comparison,rosner2002high}, as such devices represent the building blocks of many critical technologies\cite{robertson2001rfic, dicarlo2009demonstration}, including quantum information processing\cite{wallraff2004strong, raimond2001manipulating, you2011atomic, aude2017high}. For this reason, a reliable technique that gives insight on the device functioning with localized information on the device internal features - something that the usual external-port measurements cannot do\cite{bohi2010imaging} - by imaging the spatial distribution of MW fields with microscale resolution over a wide filed of view is critical for the development of the aforementioned MW technologies. 

Nitrogen-vacancy (NV) centers in diamond have emerged as a promising tool for imaging magnetic fields \cite{degenReview} over a wide range of samples and for both static and dynamic fields \cite{edlynQDMreview} due to their magnetically-sensitive fluorescence emitted upon optical excitation\cite{marcusGndState, hong2013nanoscale}. Indeed, magnetometry based on NV centers in diamond has already been employed to image static magnetic fields from current flow in 2D materials \cite{ku2020imaging, basso2022electric}, to observe ferromagnetic domains for magnetic memory \cite{simpson2016magneto} and hardware security \cite{MicroMagnets} applications, for characterization of integrated circuits \cite{kehayias2023high, Pauli555}, as well as to characterize biological samples \cite{arai2022millimetre, fescenko2019diamond}.
Imaging of MW fields has already been demonstrated, with both NV ensembles\cite{horsley2018microwave, wang2015high} and single-NV scanning probes\cite{appel2015nanoscale}, by measuring Rabi oscillations induced by the MW field within NV spin sublevels. The main limitation in these previous works is that the microwave device under test also serves as the antenna driving the NV spin transitions, which restricts the range of devices and signal strengths that can be observed. To overcome this limitation, we develop a novel method that enables to measure weak MW signals that cannot be directly measured, using a differential Rabi measurement technique.

\begin{figure*}[t]
\begin{center}
\includegraphics[width=0.9\linewidth]{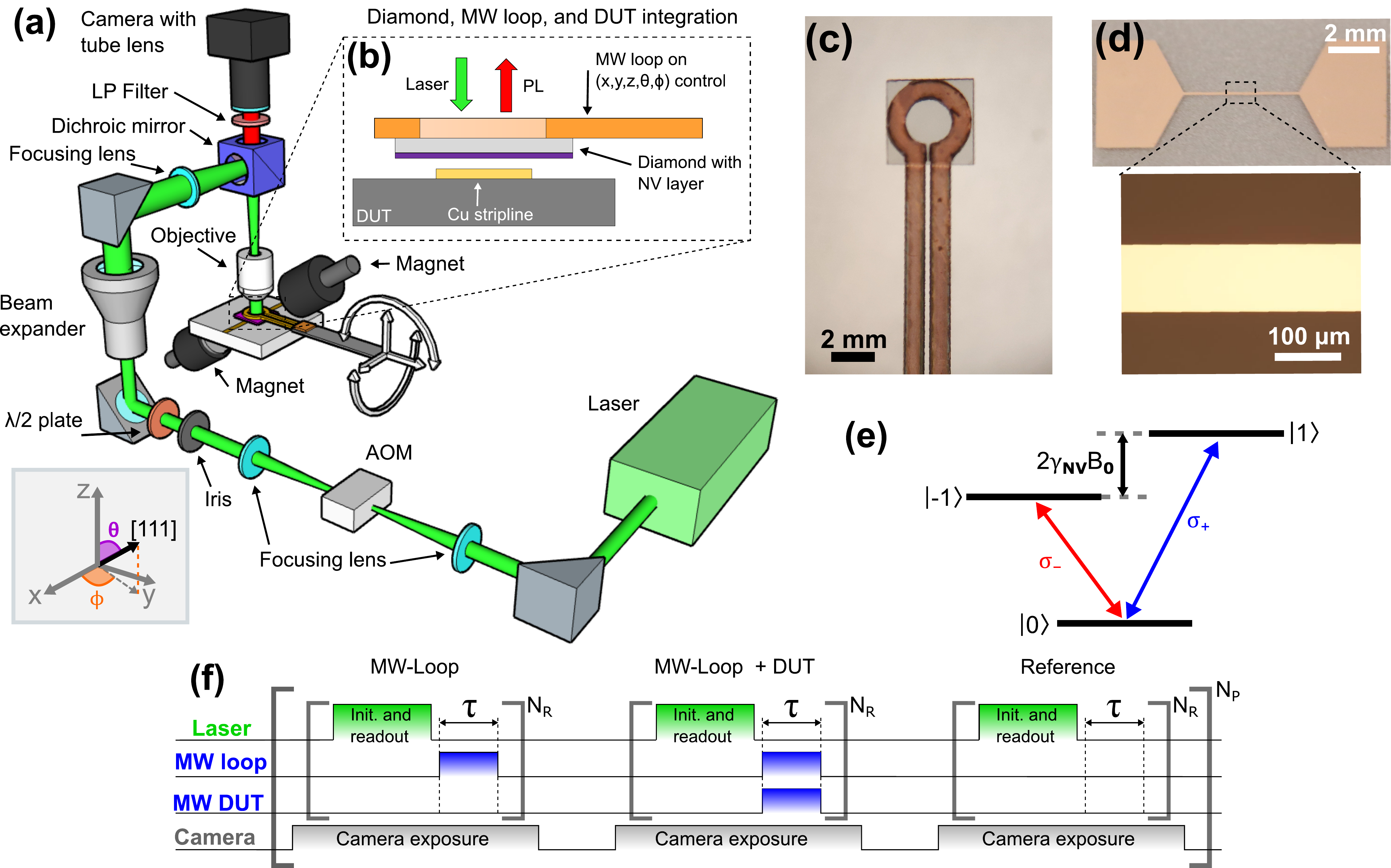}
%\captionsetup{width=1\linewidth}
\makeatletter\long\def\@ifdim#1#2#3{#2}\makeatother
\caption{(a) Schematic of the wide-field MW magnetic field imaging experimental setup. A 532 nm laser is used to excite the NV centers, while a copper MW loop delivers the MW fields to manipulate the spin states. The emitted PL is long-pass  filtered before being collected by the camera for wide field imaging. A bias magnetic field $B_{0}$ is applied along the [111]-oriented NV axis ($\theta = 55^{\circ}$, $\phi = 97^{\circ}$ in the $\left \{ x,y,z  \right \}$ laboratory reference frame) by a pair of permanent magnets. (b) Schematics of the diamond-DUT integration method. The diamond is glued on the MW loop, which is in turn mounted on a 5-axis ($x, y, z,\theta, \phi$) stage, used to bring the diamond in proximity to the DUT. (c) Optical image of the MW loop with the diamond glued on top. (d) Optical image of the DUT, consisting of a 100 $\upmu$m-wide copper wire. (e) NV center ground-state energy level structure, showing the effect of a static field $B_{0}$ lifting the $\ket{\pm 1}$ degeneracy and the $\sigma_{\pm}$-polarized MW transitions. (f) Experimental pulse sequence for wide-field Rabi oscillation measurements. Each section of the pulse sequence is repeated $N_R = 250$ times inside a single camera exposure, while the whole sequence is repeated $N_P = 100$ times for each value of MW pulse duration $\tau$.}
\label{Figure1}
\end{center}
\end{figure*}

In this work, we demonstrate wide-field imaging of MW magnetic fields in the GHz range from a device-under-test (DUT) by detecting small deviations of the NV spin sublevels Rabi oscillations around a central Rabi frequency through a differential measurement protocol. In this experiment, Rabi oscillations are driven by a MW loop operating at high power (40 dBm), positioned near the NV layer, while the DUT is deliberately operated at much lower power (in the range -2 to 28 dBm) to generate weak MW fields. We observe that, although the MW field from the DUT is too weak to directly drive Rabi oscillations, it can still be measured using a differential Rabi measurements. By employing a three-stage pulse sequence, alternating between driving Rabi oscillations with the MW loop alone and with both the MW loop and DUT, we can observe how the Rabi oscillations driven by the MW loop are altered when the DUT is active. This differential technique allows to extend the detectable MW power range by an additional $\sim$ 40 dB of DUT MW input power. The NV photoluminescence (PL) is collected with a CMOS camera, offering spatially-resolved details on the local variations in the Rabi frequency across the field of view, thus giving information on the MW field generated by the DUT. As a proof-of-principle experiment, the DUT is a fabricated 100 $\upmu$m-wide 200 nm-thick Cu stripline. After describing the experimental setup and the measurement protocol, we demonstrate the MW imaging capability of magnetic fields oscillating at 2.57 GHz over a field of view of $\sim$ $340\times340$ $\upmu$m$^2$. Moreover, we demonstrate MW imaging capabilities over dynamic range of 30 dB in the DUT input power with a measured magnetic field sensitivity of $\sim$ 9 $\upmu$T Hz$^{-1/2}$ over a single pixel with size of $\sim$ $1.2\times1.2$ $\upmu$m$^2$. In addition, to demonstrate the versatility of the Rabi differential method here developed, we image a different nontrivial DUT.  This work establishes a novel procedure to use NV quantum sensing to perform wide-field imaging of high-frequency magnetic fields. The described protocol allows us to measure weak signals, thus expanding the range of samples that can be studied by NV magnetometry, with potential applications across diverse fields, from life sciences to MW device failure analysis.

\section{Experimental methods}
\subsection{Experimental setup}
The schematics of the home-built fluorescence microscope we use to perform wide-field MW magnetic field imaging is shown in Fig. \ref{Figure1}(a). A 532 nm laser is pulsed with an AOM before being appropriately focused on the back aperture of a 20$\times$ 0.4 NA objective, to achieve a uniform illumination on a FOV of $\sim 340\times340$ $\upmu$m$^2$. The PL emitted by the NVs is then collected by a CMOS camera after being filtered with a 650 nm long-pass filter. Before analysis we apply 4x4 pixel binning, resulting in a $\sim 1.2\times 1.2$ $\upmu$m$^2$ pixel size. However, the lateral resolution in this experiment is likely limited by the NV layer thickness and the diamond-DUT stand-off distance. Two permanent magnets mounted on motorized translation stages for angle and radial control are used to apply a bias magnetic field $B_0$ aligned to the [111] crystallographic NV axis; the orientation with respect to the $\left \{ x,y,z  \right \}$ laboratory reference frame ($\theta = 55^{\circ}$, $\phi = 97^{\circ}$) is also shown in the inset of Fig. \ref{Figure1}(a). The diamond sensor, consisting of a 4 $\upmu$m-thick NV layer overgrown on a 500 $\upmu$m-thick diamond plate, is glued on the MW loop (operated at a power of 40 dBm) that is used to drive the $\ket{m_s = 0} \leftrightarrow \ket{m_s = \pm 1}$ spin transitions. The diamond/MW loop system is in turn mounted on a 5 axis ($x, y, z, \theta, \phi$) translation stage, that we use to bring the diamond in proximity of the DUT, as shown in Fig. \ref{Figure1}(b). An optical picture showing the MW loop with the diamond glued on it is reported in Fig. \ref{Figure1}(c), whereas Fig. \ref{Figure1}(d) shows an optical picture of the DUT, consisting of a 100 $\upmu$m wide, 200 nm thick Cu stripline deposited on a Al$_2$O$_3$ substrate. More details on the experimental setup are reported in the SI\cite{suppl}.

\begin{figure*}[ht]
\includegraphics[width=0.9\linewidth]{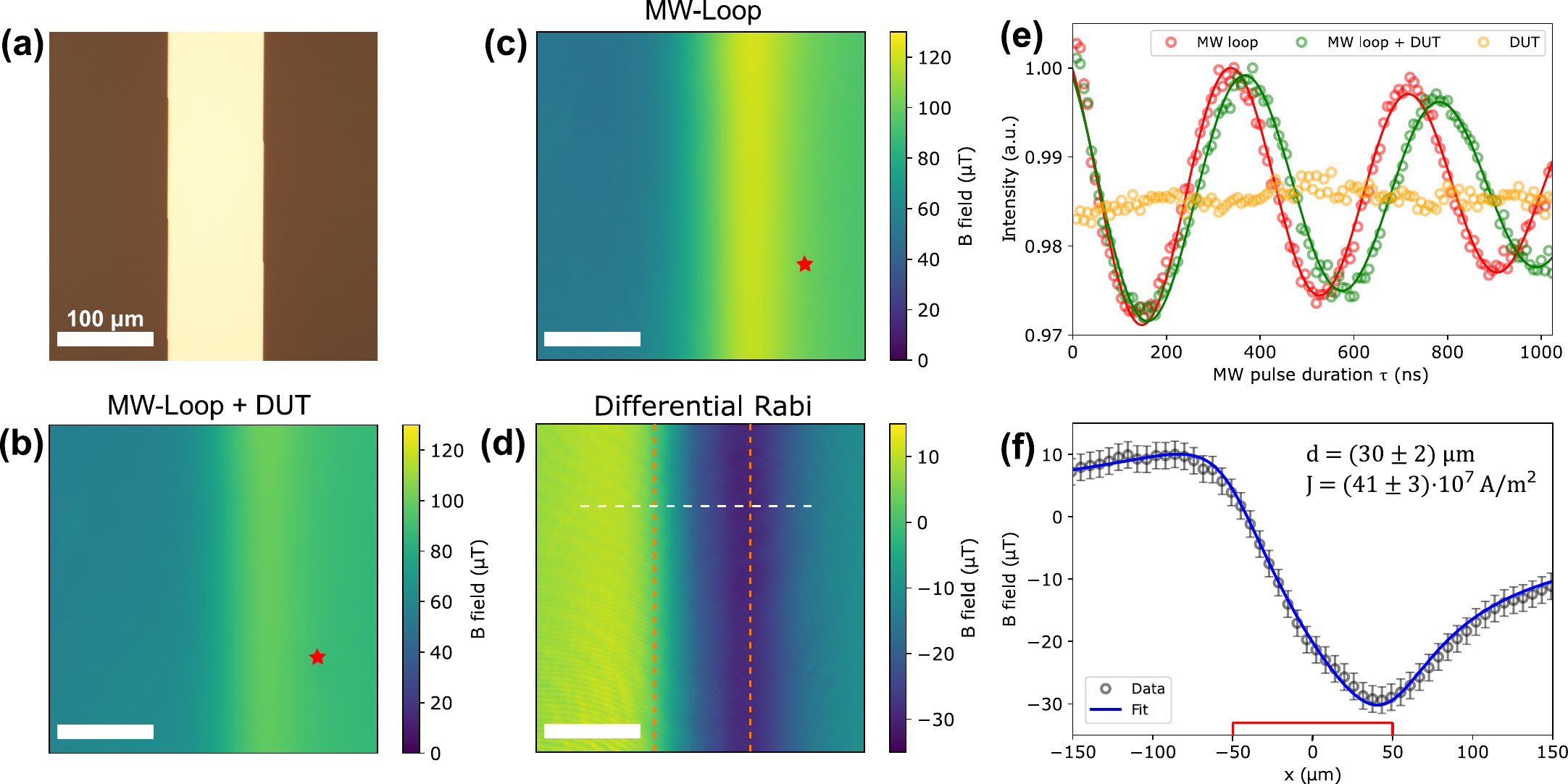}
% \captionsetup{width=1\linewidth}
\caption{(a) Optical picture of the 100 $\upmu$m-wide Cu stripline. Wide-field magnetic field imaging for the ``MW-loop + DUT'' case (b) and for the ``MW-loop'' case (c). For (b) and (c) the value of each pixel is the magnetic field obtained from $B = \Omega / (2\pi\gamma_{NV})$, where $\Omega$ is the measured Rabi frequency. (d) Results for $B_{DUT}$ obtained through the differential Rabi measurements (Eq. \ref{Eq_DUT}) The two vertical red dashed lines outline the Cu wire position. Scale bar of 100 $\upmu$m is the same for (a), (b), (c) and (d). (e) Typical single-pixel Rabi oscillations data for ``MW-loop'' (red), ``MW-loop + DUT'' (green), and ``DUT'' (yellow). Data represented here are relative to the pixel identified by the red star in (b) and (c). Circles are the experimental data while solid lines are the results of the fit obtained through Eq. \ref{Eq_fit}. (f) Magnetic field  along the horizontal white dashed line in (d). Data are fit with a model for the magnetic field generated by an uniform MW current in a stripline, using the stand-off distance $d$ and the current density $J$ as fitting parameters. Red rectangle represents the Cu wire cross section along the y axis.}
\label{Figure2}
\end{figure*}
\subsection{Detection mechanism}
The negatively-charged NV electronic ground state is a $S=1$ spin system, with the sublevels $\ket{m_s = 0}$ separated from the doubly-degenerate $\ket{m_s = \pm 1}$ sublevels by the zero-field splitting $D_{GS} \simeq 2.87$ GHz. Through the Zeeman effect, the bias magnetic field $B_0$ removes the degeneracy of the $\ket{m_s = \pm 1}$ states by a factor $2\gamma_{NV}B_0$, where $\gamma_{NV}\simeq28$ kHz $\upmu$T$^{-1}$ is the NV gyromagnetic ratio. This allows us to use either $\ket{0} \leftrightarrow \ket{+1}$ or $\ket{0} \leftrightarrow \ket{-1}$ as an isolated two-level system, as shown in Fig. \ref{Figure1}(e). Due to selection rules, the two separate transitions $\sigma_{\pm}$ ($\ket{0} \leftrightarrow \ket{\pm 1}$) are only excited by a circularly polarized MW fields\cite{appel2015nanoscale}. Thus, the two transitions $\sigma_{\pm}$ are only sensitive to a single polarization component $B_{\pm}$, with $B_{-}$ and $B_{+}$ being respectively the left- and right-handed circularly polarized component of the MW field, with the polarization axis parallel to the NV axis. A MW field resonant with either transition will then result in population oscillations between the relevant spin sublevels, i.e. Rabi oscillations, at a frequency of $\Omega_{\pm} = 2\pi\gamma_{NV}B_{\pm}$. In the following, we only use $\sigma_{-}$ for demonstration of MW magnetic field imaging. In particular we use a bias field $B_0 \sim 10.7$ mT leading to a resonant frequency $\omega_{-} = 2\pi \cdot 2.57$ GHz (measured with optically detected magnetic resonance spectrum). We emphasize that $\omega_{-}$, the frequency of the MW field we are sensitive to, can be changed simply by adjusting $B_0$. For simplicity in the notation, for the remaining of this manuscript we will suppress the ``-'' subscript as, unless otherwise stated, we are referring to only the $\sigma_{-}$ transition. 
\subsection{Pulse sequence}
We drive both the MW loop and the DUT with a MW current having same frequency $\omega = (2\pi) \cdot 2.57$ GHz and measure the Rabi oscillations with the pulse sequence reported in Fig. \ref{Figure1}(f), extended from Horsley et al~\cite{horsley2018microwave}. The sequence is divided in three sections: the first ``MW-loop'' where a MW pulse is only supplied to the MW loop line; the second ``MW-loop + DUT'' where two MW pulses of same duration are sent in both MW loop and DUT; and the third ``reference'' where the bare fluorescence of the $\ket{0}$ state is measured without applying any MW pulse. For each section, the sequence starts with a laser pulse (15 $\upmu$s duration) that both initializes the NVs into the $\ket{0}$ state and reads out their spin state. The laser pulse is followed by the MW pulse (or pulses) of duration $\tau$ that drives the Rabi oscillation. This sequence is repeated $N_R = 250$ times inside a single camera exposure time (6 ms) to accumulate PL counts. For each value of $\tau$, a total of $N_P = 100$ images are collected per each section and then averaged to further increase the signal to noise ratio. This whole sequence is repeated as we scan the MW pulse duration $\tau$ in the interval [0,1.2 $\upmu$s] to measure the Rabi oscillations. Keeping the camera open during several repetitions of the laser and MW pulse sequence leads to a decrease in the contrast, as the spin readout and the NV center repolarization to the $\ket{0}$ state occur simultaneously \cite{magaletti2024modelling}. However, this allows us to accumulate more experiment repetitions and to gate the camera at a timescale much longer than the characteristic Rabi oscillation time. At the end of the sequence we obtain three separate PL vs.~time measurements for the three MW pulse combinations described above. The ``reference'' measurement is subtracted to the other two measurements to remove the background fluorescence and correct any possible non-uniformity in the background. We then obtain two sets of measurements that we refer to respectively $B_{LOOP}$ and $B_{LOOP + DUT}$. 

\begin{figure}[t]
\centering
\includegraphics[width=1.0\linewidth]{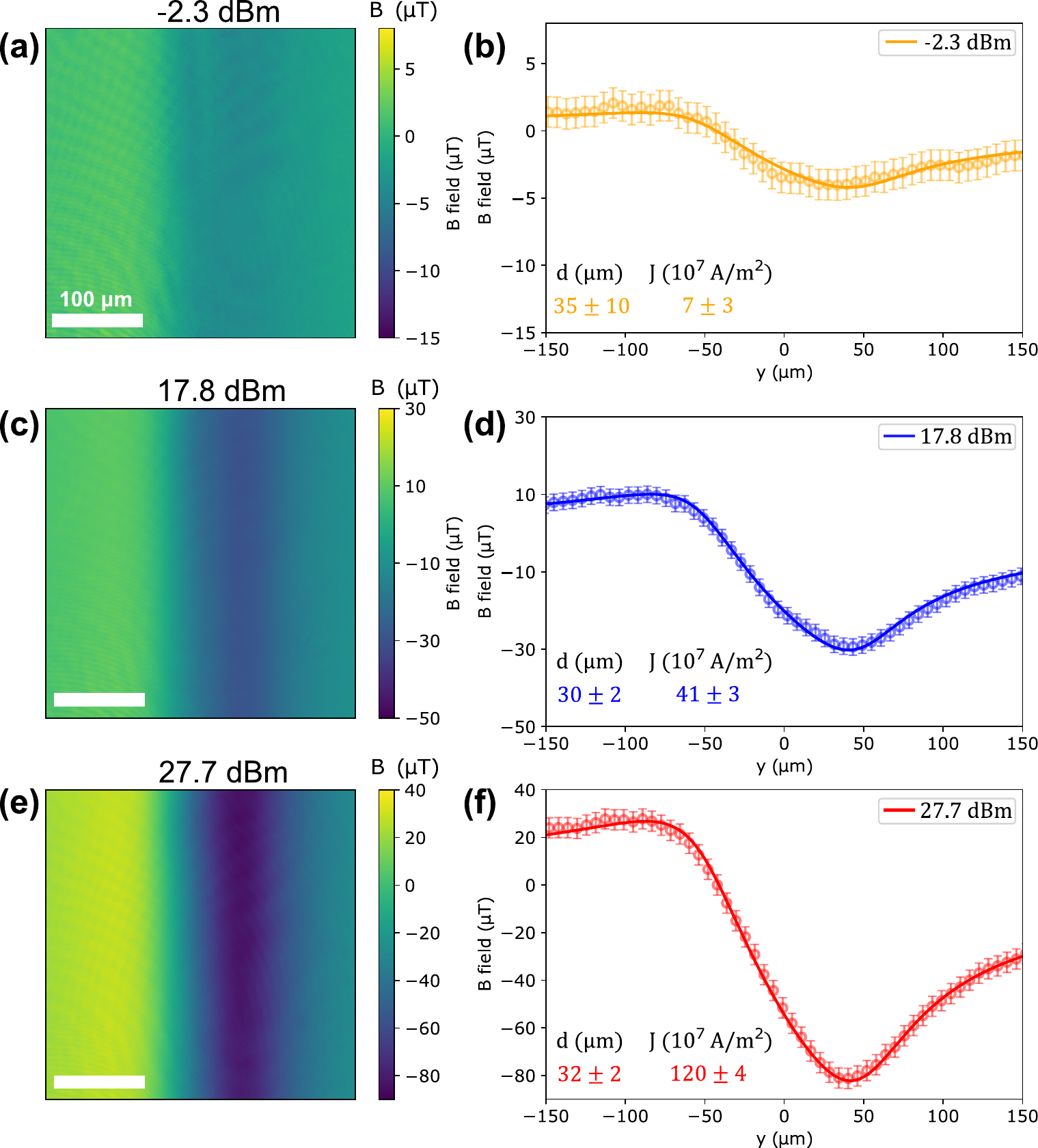}
% \captionsetup{width=1\linewidth}
\caption{Results for wide-field MW imaging obtained with the differential Rabi protocol for power of the MW injected in the DUT of -2.3 dBm (a), 17.8 dBm (c), and 27.7 dBm (e). Plots on (b), (d) and (f) show the magnetic field data along a line-cut perpendicular to the wire for, respectively, a power of -2.3 dBm, 17.8 dBm, and 27.7 dBm. The line-cut position, only shown in (a) as an horizontal white dashed lines, is the same for the three cases. The data on (b,d,f) are fit with the model for a uniform MW current in a stripline, and the results for the fitting parameters $d$ and $J$ are also reported. Scale bar of 100 $\upmu$m is the same for (a), (c), and (d).}
\label{Figure3}
\end{figure}

\section{Results and Discussion}
\subsection{MW magnetic field imaging}
Using the pulse sequence from Fig. \ref{Figure1}(f), we obtain a series of images where each pixel contains a Rabi oscillation that we fit with the equation:
\begin{equation}
    F(\Omega \tau) = F_0 + A \cos{(\Omega \tau)} \exp{(-\tau/\tau_R)} 
\label{Eq_fit}
\end{equation}
with $\tau$ being the MW pulse length while the fitting parameters are fluorescence intensity $F_0$, Rabi fluorescence contrast $A$, Rabi decay lifetime $\tau_R$, and Rabi frequency $\Omega$. We can then assign to each pixel the corresponding value of magnetic field $B$ through the relation $\Omega = 2\pi\gamma_{NV}B$. The results for the wide-field measurement obtained for a MW power in the DUT of 17.8 dBm are shown in Fig. \ref{Figure2}(b) and (c) for $B_{LOOP}$ and $B_{LOOP + DUT}$. Figure \ref{Figure2}(e) shows a typical example of Rabi oscillations for $B_{LOOP + DUT}$ (green curve) and $B_{LOOP}$ (red curve) for a single pixel (red star in Fig. \ref{Figure2}(b) and (c)). We notice that the MW field generated by the DUT is not strong enough to be detected directly. This can be observed from the yellow curve in Fig. \ref{Figure2}(e), showing a single-pixel $\tau$ scan measured when the MW pulse is sent only in the DUT (not shown in Fig. \ref{Figure1}(f)), where no Rabi oscillations can be observed. Measuring $B_{DUT}$ alone does not carry useful information, as it cannot be fit through Eq. \ref{Eq_fit}. To isolate the contribution of the DUT from the total magnetic field $\vec{B}_{Tot}(t) = \vec{B}_{LOOP}(t)+\vec{B}_{DUT}(t)$ we use the following equation
\begin{equation}
B_{DUT} = \sqrt{2} \left ( B_{LOOP + DUT} - B_{LOOP}\right)
\label{Eq_DUT}
\end{equation}
where we assumed that $|\vec{B}_{DUT}| \ll |\vec{B}_{LOOP}|$ and the $\sqrt{2}$ factor results from the root mean square (RMS) of the projection of $\vec{B}_{DUT}$ on $\vec{B}_{LOOP}$. More details on the derivation of Eq. \ref{Eq_DUT} can be found on the SI \cite{suppl}. We then take the pixel-wise difference reported in Eq. \ref{Eq_DUT} of the two images for the $B_{LOOP + DUT}$ and $B_{LOOP}$ measurements to obtain $B_{DUT}$ as shown in Fig. \ref{Figure2}(d). This result shows that our measurement protocol based on the differential measurement brings information on the MW fields generated by the DUT even if amplitude of the MW generated by the DUT alone does not drive Rabi oscillations, thus extending the range of MW power detectable through Rabi oscillations measurement.

To assess if the differential measurement for $B_{DUT}$ correctly described the MW field generated by the MW current in the DUT, we fit the MW field along a line-cut (white dashed line in Fig. \ref{Figure2}(d)) with the analytical expression that can be found for $B$ assuming an infinitely thin and long stripline carrying a uniform current density $J$ flowing only parallel to the stripline\cite{suppl}. This equation for $B$ has the stand-off distance $d$ and the DUT current density $J$ as fitting parameters. The result is shown in Fig. \ref{Figure2}(f), where a good agreement with the experimental data and the model is obtained for $d = (30\pm 2)$ $\upmu$m and $J = (41 \pm 3)\cdot 10^7$ A/m$^{2}$. The error bar on magnetic field values of Fig. \ref{Figure2}(f) are obtained starting from the standard deviation of the fit parameter $\Omega$ from Eq. \ref{Eq_fit} for the two separate measurements then propagated for Eq. \ref{Eq_DUT}.
\begin{figure}[t]
\centering
\includegraphics[width=1.0\linewidth]{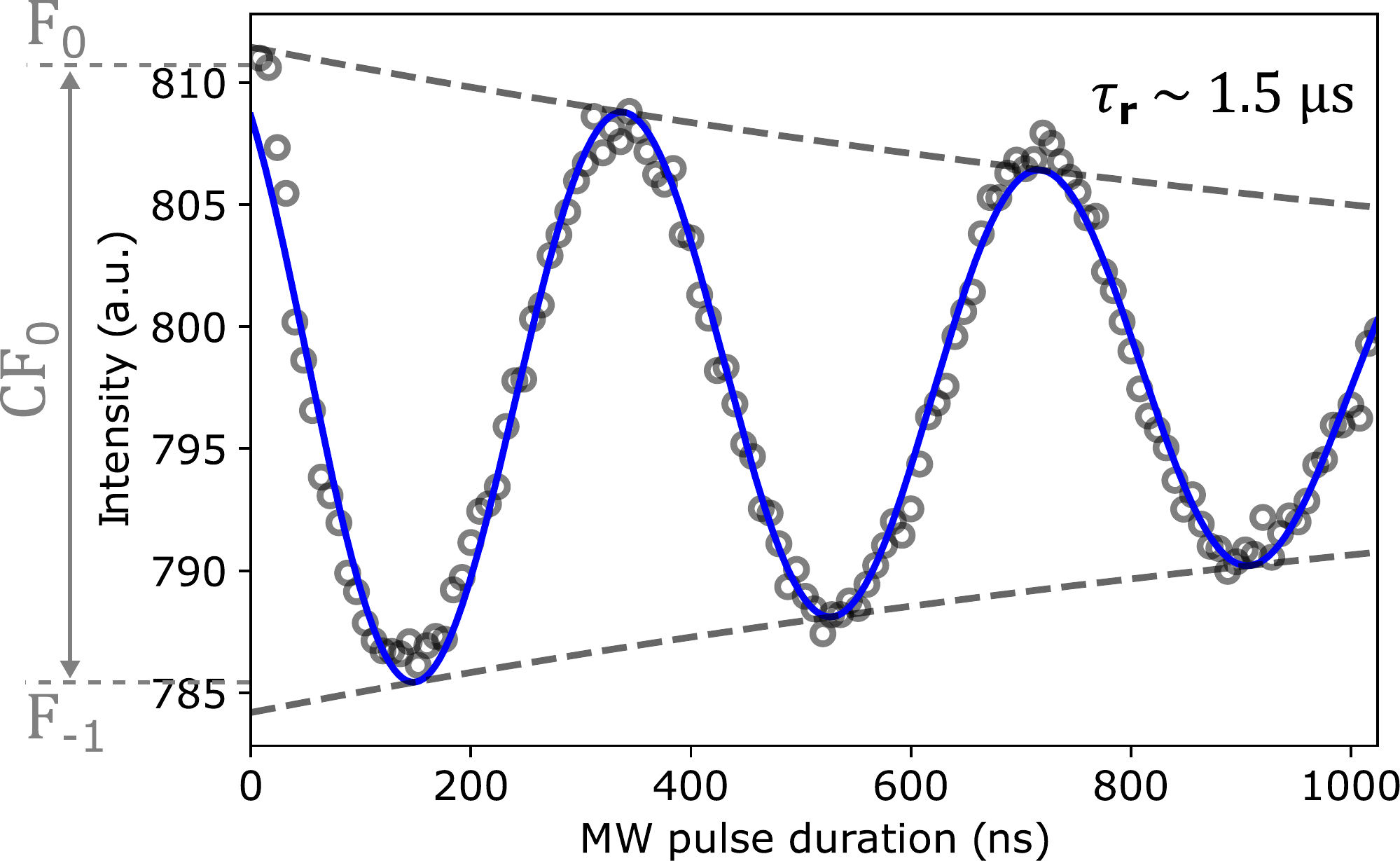}
% \captionsetup{width=1\linewidth}
\caption{Typical wide-field Rabi oscillation for the $\ket{0} \rightarrow \ket{-1}$ NV spin transition. The black circles are the experimental data while the blue line is the fit result of Eq. \ref{Eq_fit}. $\tau_R$ is the decay time of the Rabi oscillation, $F_0$ and $F_{-1}$ are the PL intensity in respectively the $\ket{0}$ and $\ket{-1}$ states, and $C = (F_0 - F_{-1})/F_0$ is the contrast.}
\label{Figure4}
\end{figure}
\subsection{Dynamic range}
To study the dynamic range accessible with the protocol developed here, we perform the same wide-field Rabi measurements for different MW power in the DUT, ranging from -2.3 dBm to 27.7 dB, with all the other same nominal experimental conditions. The results obtained through Eq. \ref{Eq_DUT} are shown in Fig. \ref{Figure3}(a), (c) and (e) for respectively -2.3, 17.8, and 27.7 dBm MW power in the DUT. Figures \ref{Figure3}(b), (d) and (f) show the magnetic field data along a line-cut perpendicular to the strip-line fit with the theoretical model. From the fitting parameters we can notice that the resulting stand-off distance $d$ is consistent in all three separate measurements. Moreover, it can be observed that the ratios between the estimated current densities $J$ are consistent with the ratios of the input power, thus proving the good quality of the magnetic field measurement. We emphasize that in this experiment, the DUT is driven with significantly weaker MW power compared to the MW loop. At higher power levels, as shown in the SI\cite{suppl}, the DUT is capable of driving Rabi oscillations on its own. However, the primary aim of this work is to develop a differential Rabi imaging technique to detect weak MW signals and show how it allows to extend the MW power range that can be detected.

\subsection{Magnetic field sensitivity}
To estimate the magnetic field sensitivity we firstly calculate the shot-noise limited sensitivity $\eta_{sn}$ through the following equation \cite{dima_magReview}:
\begin{equation}
    \eta_{SN} = \frac{\sqrt{2e}}{\pi \gamma_{NV} C \sqrt{F_0 \tau_R}}
\label{Eq_shot_noise}
\end{equation}
with $F_0$, $C$, and $\tau_R$ being respectively the fluorescence of the bright state $\ket{0}$, the contrast and the decay time of the Rabi oscillation as shown in Fig. \ref{Figure4}. For the typical values of these parameters we have in our experiment we obtain the value for the shot noise limited sensitivity of $\eta_{SN} \sim $ 670 nT Hz$^{-1/2}$. We can also estimate the measured sensitivity through the full experimental Rabi oscillation defined as $\eta_{meas} = \delta B \sqrt{T}$, where $\delta B$ is the smallest measurable magnetic field while $T$ is the total measurement time. We estimate $\delta B$ as the minimum measurable change in the Rabi frequency in $B_{LOOP + DUT}$, i.e. the standard deviation of the Rabi fit for $\Omega$. The value we extract from our data set is $\delta B \simeq 0.5$ $\upmu$T that combined with the total measurement time of $T = 357$ s lead to $\eta_{meas} \simeq 9$ $\upmu$T Hz$^{-1/2}$. As expected, this value is larger than the shot-noise limited sensitivity as it is evaluated from the measurement of a full Rabi oscillation. Eq. \ref{Eq_shot_noise} instead results from fixing the MW pulse duration relative to the point of maximal slope in the Rabi oscillation, that leads to the maximal change in fluorescence following a magnetic field variation, thus resulting in optimal sensitivity\cite{Rondin}. A better comparison can be obtained by considering that the total measurement time $T = N_{\tau}T_{single}$, where $N_{\tau}=128$ is the number of points acquired during the measurement of a full Rabi oscillation, and $T_{single}$ is the measurement time of a single point, which is limited by the camera frame rate. After normalizing by the number of points we obtain a per-point magnetic field sensitivity $\eta_{meas} \sim $ 0.7 $\upmu$T Hz$^{-1/2}$ which is comparable with the shot noise limited sensitivity.
\begin{figure}[t]
\centering
\includegraphics[width=1.0\linewidth]{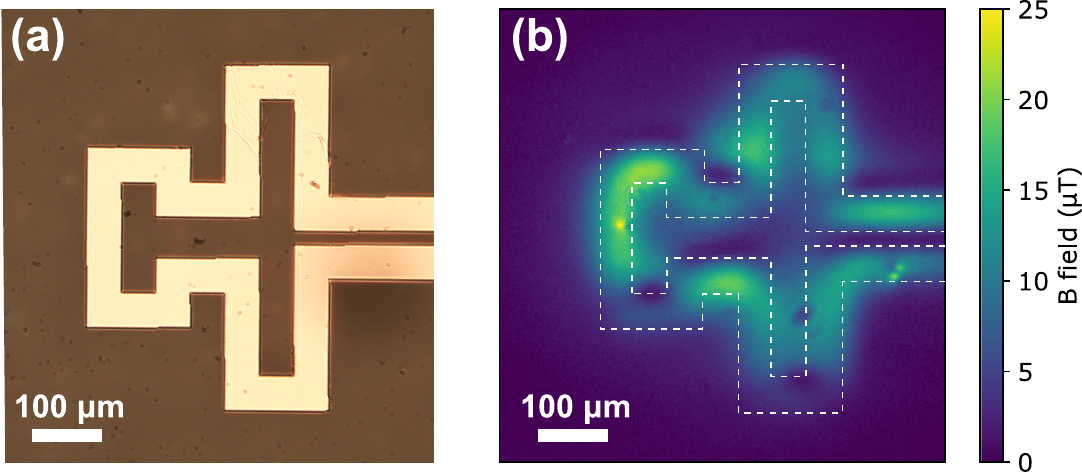}
% \captionsetup{width=1\linewidth}
\caption{(a) Optical picture of the DUT used to demonstrate MW imaging capabilities of nontrivial DUT. (b) MW magnetic field imaging through the differential Rabi measurement obtained when the DUT is driven at a MW power of 17.8 dBm.}
\label{Figure5}
\end{figure}
\subsection{MW magnetic imaging of nontrivial DUT}
In this section we demonstrate the MW imaging capabilities of the differential Rabi protocol developed in this work with another type of nontrivial DUT. The new DUT consists of a 50 $\upmu$m wide, 200 nm thick copper microstructure, deposited on a Al$_2$O$_3$ substrate in the shape shown in Fig. \ref{Figure5}(a). The result for the MW magnetic field imaging obtained with the differential Rabi protocol when the DUT is operated at 17.8 dBm is shown in Fig. \ref{Figure5}(b). 

\section{Conclusion}
In conclusion, we have demonstrated the adaptability of NV centers in diamond for wide-field imaging of high-frequency ($\simeq $ GHz) MW magnetic fields. Compared to previous results, our work expands NV-based magnetic field imaging as it is not limited to static or low-frequency fields or to image MW devices directly used to drive the NV spins transitions. With our measurement protocol based on detection of Rabi oscillations, we have successfully mapped MW magnetic field with $\upmu$m-scale spatial resolution over a $\sim$ 340×340 $\upmu$m$^2$ FOV, with a demonstrated per-pixel sensitivity of a few $\upmu$T Hz$^{-1/2}$. This work establishes a novel procedure to use NV quantum sensing to perform wide-field imaging of high-frequency magnetic fields, and demonstrates that the here developed differential Rabi imaging technique allows to gain additional range of detectable MW power. In particular, the MW field generated by the DUT can be directly measured without the aid of the differential measurement - i.e. without using the MW loop to drive the Rabi oscillations - when it is operated at a power of $\sim$ 40 dBm, as reported in the SI\cite{suppl}. The differential Rabi technique instead allows to measure the MW field generated by the DUT for input MW power down to $\sim -2$ dBm. This experiment proves that the differential Rabi imaging technique expands the range of detectable MW power by $\sim$ 40 dB. The ability we developed to image a DUT which MW field is not directly driving the NV Rabi oscillation widens the range of applicable samples and devices for MW magnetic field imaging. These results could pave the way for groundbreaking applications in NV wide-field MW fields imaging of quantum devices and materials. Future work that will further improve the impact of this technique may include for instance vector magnetometry\cite{chen2020calibration}, and a more elaborate engineering of the MW loop to achieve a uniform MW field direction and amplitude in the FOV. The latter will open the possibility of more advanced pulse sequences, such as double quantum protocols\cite{kazi2021wide, hart2021n}.

\section*{Acknowledgements}
We thank George Burns, Michael Titze, Shei Sia Su, and Edward Bielejec (Ion Beam Laboratory, Sandia National Laboratories) for help with diamond preparation.  This material is based upon work supported by the U.S. Department of Energy, Office of Science, National Quantum Information Science Research Centers, Quantum Systems Accelerator. Sandia National Laboratories is a multi-mission laboratory managed and operated by National Technology and Engineering Solutions of Sandia, LLC, a wholly owned subsidiary of Honeywell International, Inc., for the DOE's National Nuclear Security Administration under contract DE-NA0003525. This work was funded, in part, by the Laboratory Directed Research and Development Program and performed, in part, at the Center for Integrated Nanotechnologies, an Office of Science User Facility operated for the U.S.~Department of Energy (DOE) Office of Science. This paper describes objective technical results and analysis. Any subjective views or opinions that might be expressed in the paper do not necessarily represent the views of the U.S. Department of Energy or the United States Government.Additionally, this research was developed with funding from the Defense Advanced Research Projects Agency (DARPA). The views, opinions and/or findings expressed are those of the author and should not be interpreted as representing the official views or policies of the Department of Defense or the U.S. Government. Distribution Statement “A” (Approved for Public Release, Distribution Unlimited).

\section*{References}
\bibliography{Main}

%aipnum4-2.bst 2019-01-14 (MD) hand-edited version of apsrev4-1.bst
\begin{thebibliography}{35}%
\makeatletter
\providecommand \@ifxundefined [1]{%
 \@ifx{#1\undefined}
}%
\providecommand \@ifnum [1]{%
 \ifnum #1\expandafter \@firstoftwo
 \else \expandafter \@secondoftwo
 \fi
}%
\providecommand \@ifx [1]{%
 \ifx #1\expandafter \@firstoftwo
 \else \expandafter \@secondoftwo
 \fi
}%
\providecommand \natexlab [1]{#1}%
\providecommand \enquote  [1]{``#1''}%
\providecommand \bibnamefont  [1]{#1}%
\providecommand \bibfnamefont [1]{#1}%
\providecommand \citenamefont [1]{#1}%
\providecommand \href@noop [0]{\@secondoftwo}%
\providecommand \href [0]{\begingroup \@sanitize@url \@href}%
\providecommand \@href[1]{\@@startlink{#1}\@@href}%
\providecommand \@@href[1]{\endgroup#1\@@endlink}%
\providecommand \@sanitize@url [0]{\catcode `\\12\catcode `\$12\catcode `\&12\catcode `\#12\catcode `\^12\catcode `\_12\catcode `\%12\relax}%
\providecommand \@@startlink[1]{}%
\providecommand \@@endlink[0]{}%
\providecommand \url  [0]{\begingroup\@sanitize@url \@url }%
\providecommand \@url [1]{\endgroup\@href {#1}{\urlprefix }}%
\providecommand \urlprefix  [0]{URL }%
\providecommand \Eprint [0]{\href }%
\providecommand \doibase [0]{https://doi.org/}%
\providecommand \selectlanguage [0]{\@gobble}%
\providecommand \bibinfo  [0]{\@secondoftwo}%
\providecommand \bibfield  [0]{\@secondoftwo}%
\providecommand \translation [1]{[#1]}%
\providecommand \BibitemOpen [0]{}%
\providecommand \bibitemStop [0]{}%
\providecommand \bibitemNoStop [0]{.\EOS\space}%
\providecommand \EOS [0]{\spacefactor3000\relax}%
\providecommand \BibitemShut  [1]{\csname bibitem#1\endcsname}%
\let\auto@bib@innerbib\@empty
%</preamble>
\bibitem [{\citenamefont {Van~der Sar}\ \emph {et~al.}(2015)\citenamefont {Van~der Sar}, \citenamefont {Casola}, \citenamefont {Walsworth},\ and\ \citenamefont {Yacoby}}]{van2015nanometre}%
  \BibitemOpen
  \bibfield  {author} {\bibinfo {author} {\bibfnamefont {T.}~\bibnamefont {Van~der Sar}}, \bibinfo {author} {\bibfnamefont {F.}~\bibnamefont {Casola}}, \bibinfo {author} {\bibfnamefont {R.}~\bibnamefont {Walsworth}},\ and\ \bibinfo {author} {\bibfnamefont {A.}~\bibnamefont {Yacoby}},\ }\bibfield  {title} {\enquote {\bibinfo {title} {Nanometre-scale probing of spin waves using single electron spins},}\ }\href@noop {} {\bibfield  {journal} {\bibinfo  {journal} {Nature communications}\ }\textbf {\bibinfo {volume} {6}},\ \bibinfo {pages} {7886} (\bibinfo {year} {2015})}\BibitemShut {NoStop}%
\bibitem [{\citenamefont {Balents}(2010)}]{balents2010spin}%
  \BibitemOpen
  \bibfield  {author} {\bibinfo {author} {\bibfnamefont {L.}~\bibnamefont {Balents}},\ }\bibfield  {title} {\enquote {\bibinfo {title} {Spin liquids in frustrated magnets},}\ }\href@noop {} {\bibfield  {journal} {\bibinfo  {journal} {nature}\ }\textbf {\bibinfo {volume} {464}},\ \bibinfo {pages} {199--208} (\bibinfo {year} {2010})}\BibitemShut {NoStop}%
\bibitem [{\citenamefont {Dasika}, \citenamefont {Parashar},\ and\ \citenamefont {Saha}(2023)}]{dasika2023mapping}%
  \BibitemOpen
  \bibfield  {author} {\bibinfo {author} {\bibfnamefont {S.}~\bibnamefont {Dasika}}, \bibinfo {author} {\bibfnamefont {M.}~\bibnamefont {Parashar}},\ and\ \bibinfo {author} {\bibfnamefont {K.}~\bibnamefont {Saha}},\ }\bibfield  {title} {\enquote {\bibinfo {title} {Mapping ac susceptibility with quantum diamond microscope},}\ }\href@noop {} {\bibfield  {journal} {\bibinfo  {journal} {Rev. Sci. Instrum.}\ }\textbf {\bibinfo {volume} {94}} (\bibinfo {year} {2023})}\BibitemShut {NoStop}%
\bibitem [{\citenamefont {Hall}, \citenamefont {Simpson},\ and\ \citenamefont {Hollenberg}(2013)}]{hall2013nanoscale}%
  \BibitemOpen
  \bibfield  {author} {\bibinfo {author} {\bibfnamefont {L.}~\bibnamefont {Hall}}, \bibinfo {author} {\bibfnamefont {D.}~\bibnamefont {Simpson}},\ and\ \bibinfo {author} {\bibfnamefont {L.}~\bibnamefont {Hollenberg}},\ }\bibfield  {title} {\enquote {\bibinfo {title} {Nanoscale sensing and imaging in biology using the nitrogen-vacancy center in diamond},}\ }\href@noop {} {\bibfield  {journal} {\bibinfo  {journal} {Mrs Bulletin}\ }\textbf {\bibinfo {volume} {38}},\ \bibinfo {pages} {162--167} (\bibinfo {year} {2013})}\BibitemShut {NoStop}%
\bibitem [{\citenamefont {Sayil}, \citenamefont {Kerns},\ and\ \citenamefont {Kerns}(2005)}]{sayil2005comparison}%
  \BibitemOpen
  \bibfield  {author} {\bibinfo {author} {\bibfnamefont {S.}~\bibnamefont {Sayil}}, \bibinfo {author} {\bibfnamefont {D.~V.}\ \bibnamefont {Kerns}},\ and\ \bibinfo {author} {\bibfnamefont {S.~E.}\ \bibnamefont {Kerns}},\ }\bibfield  {title} {\enquote {\bibinfo {title} {Comparison of contactless measurement and testing techniques to a all-silicon optical test and characterization method},}\ }\href@noop {} {\bibfield  {journal} {\bibinfo  {journal} {IEEE Transactions on Instrumentation and Measurement}\ }\textbf {\bibinfo {volume} {54}},\ \bibinfo {pages} {2082--2089} (\bibinfo {year} {2005})}\BibitemShut {NoStop}%
\bibitem [{\citenamefont {Rosner}\ and\ \citenamefont {Van Der~Weide}(2002)}]{rosner2002high}%
  \BibitemOpen
  \bibfield  {author} {\bibinfo {author} {\bibfnamefont {B.~T.}\ \bibnamefont {Rosner}}\ and\ \bibinfo {author} {\bibfnamefont {D.~W.}\ \bibnamefont {Van Der~Weide}},\ }\bibfield  {title} {\enquote {\bibinfo {title} {High-frequency near-field microscopy},}\ }\href@noop {} {\bibfield  {journal} {\bibinfo  {journal} {Review of Scientific Instruments}\ }\textbf {\bibinfo {volume} {73}},\ \bibinfo {pages} {2505--2525} (\bibinfo {year} {2002})}\BibitemShut {NoStop}%
\bibitem [{\citenamefont {Robertson}\ and\ \citenamefont {Lucyszyn}(2001)}]{robertson2001rfic}%
  \BibitemOpen
  \bibfield  {author} {\bibinfo {author} {\bibfnamefont {I.~D.}\ \bibnamefont {Robertson}}\ and\ \bibinfo {author} {\bibfnamefont {S.}~\bibnamefont {Lucyszyn}},\ }\href@noop {} {\emph {\bibinfo {title} {RFIC and MMIC Design and Technology}}},\ \bibinfo {number} {13}\ (\bibinfo  {publisher} {Iet},\ \bibinfo {year} {2001})\BibitemShut {NoStop}%
\bibitem [{\citenamefont {DiCarlo}\ \emph {et~al.}(2009)\citenamefont {DiCarlo}, \citenamefont {Chow}, \citenamefont {Gambetta}, \citenamefont {Bishop}, \citenamefont {Johnson}, \citenamefont {Schuster}, \citenamefont {Majer}, \citenamefont {Blais}, \citenamefont {Frunzio}, \citenamefont {Girvin} \emph {et~al.}}]{dicarlo2009demonstration}%
  \BibitemOpen
  \bibfield  {author} {\bibinfo {author} {\bibfnamefont {L.}~\bibnamefont {DiCarlo}}, \bibinfo {author} {\bibfnamefont {J.~M.}\ \bibnamefont {Chow}}, \bibinfo {author} {\bibfnamefont {J.~M.}\ \bibnamefont {Gambetta}}, \bibinfo {author} {\bibfnamefont {L.~S.}\ \bibnamefont {Bishop}}, \bibinfo {author} {\bibfnamefont {B.~R.}\ \bibnamefont {Johnson}}, \bibinfo {author} {\bibfnamefont {D.}~\bibnamefont {Schuster}}, \bibinfo {author} {\bibfnamefont {J.}~\bibnamefont {Majer}}, \bibinfo {author} {\bibfnamefont {A.}~\bibnamefont {Blais}}, \bibinfo {author} {\bibfnamefont {L.}~\bibnamefont {Frunzio}}, \bibinfo {author} {\bibfnamefont {S.}~\bibnamefont {Girvin}}, \emph {et~al.},\ }\bibfield  {title} {\enquote {\bibinfo {title} {Demonstration of two-qubit algorithms with a superconducting quantum processor},}\ }\href@noop {} {\bibfield  {journal} {\bibinfo  {journal} {Nature}\ }\textbf {\bibinfo {volume} {460}},\ \bibinfo {pages} {240--244} (\bibinfo {year} {2009})}\BibitemShut {NoStop}%
\bibitem [{\citenamefont {Wallraff}\ \emph {et~al.}(2004)\citenamefont {Wallraff}, \citenamefont {Schuster}, \citenamefont {Blais}, \citenamefont {Frunzio}, \citenamefont {Huang}, \citenamefont {Majer}, \citenamefont {Kumar}, \citenamefont {Girvin},\ and\ \citenamefont {Schoelkopf}}]{wallraff2004strong}%
  \BibitemOpen
  \bibfield  {author} {\bibinfo {author} {\bibfnamefont {A.}~\bibnamefont {Wallraff}}, \bibinfo {author} {\bibfnamefont {D.~I.}\ \bibnamefont {Schuster}}, \bibinfo {author} {\bibfnamefont {A.}~\bibnamefont {Blais}}, \bibinfo {author} {\bibfnamefont {L.}~\bibnamefont {Frunzio}}, \bibinfo {author} {\bibfnamefont {R.-S.}\ \bibnamefont {Huang}}, \bibinfo {author} {\bibfnamefont {J.}~\bibnamefont {Majer}}, \bibinfo {author} {\bibfnamefont {S.}~\bibnamefont {Kumar}}, \bibinfo {author} {\bibfnamefont {S.~M.}\ \bibnamefont {Girvin}},\ and\ \bibinfo {author} {\bibfnamefont {R.~J.}\ \bibnamefont {Schoelkopf}},\ }\bibfield  {title} {\enquote {\bibinfo {title} {Strong coupling of a single photon to a superconducting qubit using circuit quantum electrodynamics},}\ }\href@noop {} {\bibfield  {journal} {\bibinfo  {journal} {Nature}\ }\textbf {\bibinfo {volume} {431}},\ \bibinfo {pages} {162--167} (\bibinfo {year} {2004})}\BibitemShut {NoStop}%
\bibitem [{\citenamefont {Raimond}, \citenamefont {Brune},\ and\ \citenamefont {Haroche}(2001)}]{raimond2001manipulating}%
  \BibitemOpen
  \bibfield  {author} {\bibinfo {author} {\bibfnamefont {J.-M.}\ \bibnamefont {Raimond}}, \bibinfo {author} {\bibfnamefont {M.}~\bibnamefont {Brune}},\ and\ \bibinfo {author} {\bibfnamefont {S.}~\bibnamefont {Haroche}},\ }\bibfield  {title} {\enquote {\bibinfo {title} {Manipulating quantum entanglement with atoms and photons in a cavity},}\ }\href@noop {} {\bibfield  {journal} {\bibinfo  {journal} {Reviews of Modern Physics}\ }\textbf {\bibinfo {volume} {73}},\ \bibinfo {pages} {565} (\bibinfo {year} {2001})}\BibitemShut {NoStop}%
\bibitem [{\citenamefont {You}\ and\ \citenamefont {Nori}(2011)}]{you2011atomic}%
  \BibitemOpen
  \bibfield  {author} {\bibinfo {author} {\bibfnamefont {J.-Q.}\ \bibnamefont {You}}\ and\ \bibinfo {author} {\bibfnamefont {F.}~\bibnamefont {Nori}},\ }\bibfield  {title} {\enquote {\bibinfo {title} {Atomic physics and quantum optics using superconducting circuits},}\ }\href@noop {} {\bibfield  {journal} {\bibinfo  {journal} {Nature}\ }\textbf {\bibinfo {volume} {474}},\ \bibinfo {pages} {589--597} (\bibinfo {year} {2011})}\BibitemShut {NoStop}%
\bibitem [{\citenamefont {Aude~Craik}\ \emph {et~al.}(2017)\citenamefont {Aude~Craik}, \citenamefont {Linke}, \citenamefont {Sepiol}, \citenamefont {Harty}, \citenamefont {Goodwin}, \citenamefont {Ballance}, \citenamefont {Stacey}, \citenamefont {Steane}, \citenamefont {Lucas},\ and\ \citenamefont {Allcock}}]{aude2017high}%
  \BibitemOpen
  \bibfield  {author} {\bibinfo {author} {\bibfnamefont {D.}~\bibnamefont {Aude~Craik}}, \bibinfo {author} {\bibfnamefont {N.}~\bibnamefont {Linke}}, \bibinfo {author} {\bibfnamefont {M.}~\bibnamefont {Sepiol}}, \bibinfo {author} {\bibfnamefont {T.}~\bibnamefont {Harty}}, \bibinfo {author} {\bibfnamefont {J.}~\bibnamefont {Goodwin}}, \bibinfo {author} {\bibfnamefont {C.}~\bibnamefont {Ballance}}, \bibinfo {author} {\bibfnamefont {D.}~\bibnamefont {Stacey}}, \bibinfo {author} {\bibfnamefont {A.}~\bibnamefont {Steane}}, \bibinfo {author} {\bibfnamefont {D.}~\bibnamefont {Lucas}},\ and\ \bibinfo {author} {\bibfnamefont {D.}~\bibnamefont {Allcock}},\ }\bibfield  {title} {\enquote {\bibinfo {title} {High-fidelity spatial and polarization addressing of ca+ 43 qubits using near-field microwave control},}\ }\href@noop {} {\bibfield  {journal} {\bibinfo  {journal} {Phys. Rev. A}\ }\textbf {\bibinfo {volume} {95}},\ \bibinfo {pages} {022337} (\bibinfo {year} {2017})}\BibitemShut {NoStop}%
\bibitem [{\citenamefont {B{\"o}hi}\ \emph {et~al.}(2010)\citenamefont {B{\"o}hi}, \citenamefont {Riedel}, \citenamefont {H{\"a}nsch},\ and\ \citenamefont {Treutlein}}]{bohi2010imaging}%
  \BibitemOpen
  \bibfield  {author} {\bibinfo {author} {\bibfnamefont {P.}~\bibnamefont {B{\"o}hi}}, \bibinfo {author} {\bibfnamefont {M.~F.}\ \bibnamefont {Riedel}}, \bibinfo {author} {\bibfnamefont {T.~W.}\ \bibnamefont {H{\"a}nsch}},\ and\ \bibinfo {author} {\bibfnamefont {P.}~\bibnamefont {Treutlein}},\ }\bibfield  {title} {\enquote {\bibinfo {title} {Imaging of microwave fields using ultracold atoms},}\ }\href@noop {} {\bibfield  {journal} {\bibinfo  {journal} {Applied Physics Letters}\ }\textbf {\bibinfo {volume} {97}} (\bibinfo {year} {2010})}\BibitemShut {NoStop}%
\bibitem [{\citenamefont {Schirhagl}\ \emph {et~al.}(2014)\citenamefont {Schirhagl}, \citenamefont {Chang}, \citenamefont {Loretz},\ and\ \citenamefont {Degen}}]{degenReview}%
  \BibitemOpen
  \bibfield  {author} {\bibinfo {author} {\bibfnamefont {R.}~\bibnamefont {Schirhagl}}, \bibinfo {author} {\bibfnamefont {K.}~\bibnamefont {Chang}}, \bibinfo {author} {\bibfnamefont {M.}~\bibnamefont {Loretz}},\ and\ \bibinfo {author} {\bibfnamefont {C.~L.}\ \bibnamefont {Degen}},\ }\bibfield  {title} {\enquote {\bibinfo {title} {Nitrogen-vacancy centers in diamond: Nanoscale sensors for physics and biology},}\ }\href {https://doi.org/10.1146/annurev-physchem-040513-103659} {\bibfield  {journal} {\bibinfo  {journal} {Annual Review of Physical Chemistry}\ }\textbf {\bibinfo {volume} {65}},\ \bibinfo {pages} {83--105} (\bibinfo {year} {2014})}\BibitemShut {NoStop}%
\bibitem [{\citenamefont {Levine}\ \emph {et~al.}(2019)\citenamefont {Levine}, \citenamefont {Turner}, \citenamefont {Kehayias}, \citenamefont {Hart}, \citenamefont {Langellier}, \citenamefont {Trubko}, \citenamefont {Glenn}, \citenamefont {Fu},\ and\ \citenamefont {Walsworth}}]{edlynQDMreview}%
  \BibitemOpen
  \bibfield  {author} {\bibinfo {author} {\bibfnamefont {E.~V.}\ \bibnamefont {Levine}}, \bibinfo {author} {\bibfnamefont {M.~J.}\ \bibnamefont {Turner}}, \bibinfo {author} {\bibfnamefont {P.}~\bibnamefont {Kehayias}}, \bibinfo {author} {\bibfnamefont {C.~A.}\ \bibnamefont {Hart}}, \bibinfo {author} {\bibfnamefont {N.}~\bibnamefont {Langellier}}, \bibinfo {author} {\bibfnamefont {R.}~\bibnamefont {Trubko}}, \bibinfo {author} {\bibfnamefont {D.~R.}\ \bibnamefont {Glenn}}, \bibinfo {author} {\bibfnamefont {R.~R.}\ \bibnamefont {Fu}},\ and\ \bibinfo {author} {\bibfnamefont {R.~L.}\ \bibnamefont {Walsworth}},\ }\bibfield  {title} {\enquote {\bibinfo {title} {Principles and techniques of the quantum diamond microscope},}\ }\href@noop {} {\bibfield  {journal} {\bibinfo  {journal} {Nanophotonics}\ }\textbf {\bibinfo {volume} {8}},\ \bibinfo {pages} {1945--1973} (\bibinfo {year} {2019})}\BibitemShut {NoStop}%
\bibitem [{\citenamefont {Doherty}\ \emph {et~al.}(2012)\citenamefont {Doherty}, \citenamefont {Dolde}, \citenamefont {Fedder}, \citenamefont {Jelezko}, \citenamefont {Wrachtrup}, \citenamefont {Manson},\ and\ \citenamefont {Hollenberg}}]{marcusGndState}%
  \BibitemOpen
  \bibfield  {author} {\bibinfo {author} {\bibfnamefont {M.~W.}\ \bibnamefont {Doherty}}, \bibinfo {author} {\bibfnamefont {F.}~\bibnamefont {Dolde}}, \bibinfo {author} {\bibfnamefont {H.}~\bibnamefont {Fedder}}, \bibinfo {author} {\bibfnamefont {F.}~\bibnamefont {Jelezko}}, \bibinfo {author} {\bibfnamefont {J.}~\bibnamefont {Wrachtrup}}, \bibinfo {author} {\bibfnamefont {N.~B.}\ \bibnamefont {Manson}},\ and\ \bibinfo {author} {\bibfnamefont {L.~C.~L.}\ \bibnamefont {Hollenberg}},\ }\bibfield  {title} {\enquote {\bibinfo {title} {Theory of the ground-state spin of the nv$^-$ center in diamond},}\ }\href {https://doi.org/10.1103/PhysRevB.85.205203} {\bibfield  {journal} {\bibinfo  {journal} {Phys. Rev. B}\ }\textbf {\bibinfo {volume} {85}},\ \bibinfo {pages} {205203} (\bibinfo {year} {2012})}\BibitemShut {NoStop}%
\bibitem [{\citenamefont {Hong}\ \emph {et~al.}(2013)\citenamefont {Hong}, \citenamefont {Grinolds}, \citenamefont {Pham}, \citenamefont {Le~Sage}, \citenamefont {Luan}, \citenamefont {Walsworth},\ and\ \citenamefont {Yacoby}}]{hong2013nanoscale}%
  \BibitemOpen
  \bibfield  {author} {\bibinfo {author} {\bibfnamefont {S.}~\bibnamefont {Hong}}, \bibinfo {author} {\bibfnamefont {M.~S.}\ \bibnamefont {Grinolds}}, \bibinfo {author} {\bibfnamefont {L.~M.}\ \bibnamefont {Pham}}, \bibinfo {author} {\bibfnamefont {D.}~\bibnamefont {Le~Sage}}, \bibinfo {author} {\bibfnamefont {L.}~\bibnamefont {Luan}}, \bibinfo {author} {\bibfnamefont {R.~L.}\ \bibnamefont {Walsworth}},\ and\ \bibinfo {author} {\bibfnamefont {A.}~\bibnamefont {Yacoby}},\ }\bibfield  {title} {\enquote {\bibinfo {title} {Nanoscale magnetometry with nv centers in diamond},}\ }\href@noop {} {\bibfield  {journal} {\bibinfo  {journal} {MRS bulletin}\ }\textbf {\bibinfo {volume} {38}},\ \bibinfo {pages} {155--161} (\bibinfo {year} {2013})}\BibitemShut {NoStop}%
\bibitem [{\citenamefont {Ku}\ \emph {et~al.}(2020)\citenamefont {Ku}, \citenamefont {Zhou}, \citenamefont {Li}, \citenamefont {Shin}, \citenamefont {Shi}, \citenamefont {Burch}, \citenamefont {Anderson}, \citenamefont {Pierce}, \citenamefont {Xie}, \citenamefont {Hamo} \emph {et~al.}}]{ku2020imaging}%
  \BibitemOpen
  \bibfield  {author} {\bibinfo {author} {\bibfnamefont {M.~J.}\ \bibnamefont {Ku}}, \bibinfo {author} {\bibfnamefont {T.~X.}\ \bibnamefont {Zhou}}, \bibinfo {author} {\bibfnamefont {Q.}~\bibnamefont {Li}}, \bibinfo {author} {\bibfnamefont {Y.~J.}\ \bibnamefont {Shin}}, \bibinfo {author} {\bibfnamefont {J.~K.}\ \bibnamefont {Shi}}, \bibinfo {author} {\bibfnamefont {C.}~\bibnamefont {Burch}}, \bibinfo {author} {\bibfnamefont {L.~E.}\ \bibnamefont {Anderson}}, \bibinfo {author} {\bibfnamefont {A.~T.}\ \bibnamefont {Pierce}}, \bibinfo {author} {\bibfnamefont {Y.}~\bibnamefont {Xie}}, \bibinfo {author} {\bibfnamefont {A.}~\bibnamefont {Hamo}}, \emph {et~al.},\ }\bibfield  {title} {\enquote {\bibinfo {title} {Imaging viscous flow of the dirac fluid in graphene},}\ }\href@noop {} {\bibfield  {journal} {\bibinfo  {journal} {Nature}\ }\textbf {\bibinfo {volume} {583}},\ \bibinfo {pages} {537--541} (\bibinfo {year} {2020})}\BibitemShut {NoStop}%
\bibitem [{\citenamefont {Basso}\ \emph {et~al.}(2022)\citenamefont {Basso}, \citenamefont {Kehayias}, \citenamefont {Henshaw}, \citenamefont {Ziabari}, \citenamefont {Byeon}, \citenamefont {Lilly}, \citenamefont {Bussmann}, \citenamefont {Campbell}, \citenamefont {Misra},\ and\ \citenamefont {Mounce}}]{basso2022electric}%
  \BibitemOpen
  \bibfield  {author} {\bibinfo {author} {\bibfnamefont {L.}~\bibnamefont {Basso}}, \bibinfo {author} {\bibfnamefont {P.}~\bibnamefont {Kehayias}}, \bibinfo {author} {\bibfnamefont {J.}~\bibnamefont {Henshaw}}, \bibinfo {author} {\bibfnamefont {M.~S.}\ \bibnamefont {Ziabari}}, \bibinfo {author} {\bibfnamefont {H.}~\bibnamefont {Byeon}}, \bibinfo {author} {\bibfnamefont {M.~P.}\ \bibnamefont {Lilly}}, \bibinfo {author} {\bibfnamefont {E.}~\bibnamefont {Bussmann}}, \bibinfo {author} {\bibfnamefont {D.~M.}\ \bibnamefont {Campbell}}, \bibinfo {author} {\bibfnamefont {S.}~\bibnamefont {Misra}},\ and\ \bibinfo {author} {\bibfnamefont {A.~M.}\ \bibnamefont {Mounce}},\ }\bibfield  {title} {\enquote {\bibinfo {title} {Electric current paths in a si: P delta-doped device imaged by nitrogen-vacancy diamond magnetic microscopy},}\ }\href@noop {} {\bibfield  {journal} {\bibinfo  {journal} {Nanotechnology}\ }\textbf {\bibinfo {volume} {34}},\ \bibinfo {pages} {015001} (\bibinfo {year} {2022})}\BibitemShut {NoStop}%
\bibitem [{\citenamefont {Simpson}\ \emph {et~al.}(2016)\citenamefont {Simpson}, \citenamefont {Tetienne}, \citenamefont {McCoey}, \citenamefont {Ganesan}, \citenamefont {Hall}, \citenamefont {Petrou}, \citenamefont {Scholten},\ and\ \citenamefont {Hollenberg}}]{simpson2016magneto}%
  \BibitemOpen
  \bibfield  {author} {\bibinfo {author} {\bibfnamefont {D.~A.}\ \bibnamefont {Simpson}}, \bibinfo {author} {\bibfnamefont {J.-P.}\ \bibnamefont {Tetienne}}, \bibinfo {author} {\bibfnamefont {J.~M.}\ \bibnamefont {McCoey}}, \bibinfo {author} {\bibfnamefont {K.}~\bibnamefont {Ganesan}}, \bibinfo {author} {\bibfnamefont {L.~T.}\ \bibnamefont {Hall}}, \bibinfo {author} {\bibfnamefont {S.}~\bibnamefont {Petrou}}, \bibinfo {author} {\bibfnamefont {R.~E.}\ \bibnamefont {Scholten}},\ and\ \bibinfo {author} {\bibfnamefont {L.~C.}\ \bibnamefont {Hollenberg}},\ }\bibfield  {title} {\enquote {\bibinfo {title} {Magneto-optical imaging of thin magnetic films using spins in diamond},}\ }\href@noop {} {\bibfield  {journal} {\bibinfo  {journal} {Scientific reports}\ }\textbf {\bibinfo {volume} {6}},\ \bibinfo {pages} {22797} (\bibinfo {year} {2016})}\BibitemShut {NoStop}%
\bibitem [{\citenamefont {Kehayias}\ \emph {et~al.}(2020)\citenamefont {Kehayias}, \citenamefont {Bussmann}, \citenamefont {Lu},\ and\ \citenamefont {Mounce}}]{MicroMagnets}%
  \BibitemOpen
  \bibfield  {author} {\bibinfo {author} {\bibfnamefont {P.}~\bibnamefont {Kehayias}}, \bibinfo {author} {\bibfnamefont {E.}~\bibnamefont {Bussmann}}, \bibinfo {author} {\bibfnamefont {T.~M.}\ \bibnamefont {Lu}},\ and\ \bibinfo {author} {\bibfnamefont {A.~M.}\ \bibnamefont {Mounce}},\ }\bibfield  {title} {\enquote {\bibinfo {title} {A physically unclonable function using nv diamond magnetometry and micromagnet arrays},}\ }\href@noop {} {\bibfield  {journal} {\bibinfo  {journal} {J. Appl. Phys.}\ }\textbf {\bibinfo {volume} {127}},\ \bibinfo {pages} {203904} (\bibinfo {year} {2020})}\BibitemShut {NoStop}%
\bibitem [{\citenamefont {Kehayias}\ \emph {et~al.}(2023)\citenamefont {Kehayias}, \citenamefont {Walraven}, \citenamefont {Rodarte},\ and\ \citenamefont {Mounce}}]{kehayias2023high}%
  \BibitemOpen
  \bibfield  {author} {\bibinfo {author} {\bibfnamefont {P.}~\bibnamefont {Kehayias}}, \bibinfo {author} {\bibfnamefont {J.}~\bibnamefont {Walraven}}, \bibinfo {author} {\bibfnamefont {A.}~\bibnamefont {Rodarte}},\ and\ \bibinfo {author} {\bibfnamefont {A.}~\bibnamefont {Mounce}},\ }\bibfield  {title} {\enquote {\bibinfo {title} {High-resolution short-circuit fault localization in a multilayer integrated circuit using a quantum diamond microscope},}\ }\href@noop {} {\bibfield  {journal} {\bibinfo  {journal} {Physical Review Applied}\ }\textbf {\bibinfo {volume} {20}},\ \bibinfo {pages} {014036} (\bibinfo {year} {2023})}\BibitemShut {NoStop}%
\bibitem [{\citenamefont {Kehayias}\ \emph {et~al.}(2022)\citenamefont {Kehayias}, \citenamefont {Levine}, \citenamefont {Basso}, \citenamefont {Henshaw}, \citenamefont {Saleh~Ziabari}, \citenamefont {Titze}, \citenamefont {Haltli}, \citenamefont {Okoro}, \citenamefont {Tibbetts}, \citenamefont {Udoni}, \citenamefont {Bielejec}, \citenamefont {Lilly}, \citenamefont {Lu}, \citenamefont {Schwindt},\ and\ \citenamefont {Mounce}}]{Pauli555}%
  \BibitemOpen
  \bibfield  {author} {\bibinfo {author} {\bibfnamefont {P.}~\bibnamefont {Kehayias}}, \bibinfo {author} {\bibfnamefont {E.~V.}\ \bibnamefont {Levine}}, \bibinfo {author} {\bibfnamefont {L.}~\bibnamefont {Basso}}, \bibinfo {author} {\bibfnamefont {J.}~\bibnamefont {Henshaw}}, \bibinfo {author} {\bibfnamefont {M.}~\bibnamefont {Saleh~Ziabari}}, \bibinfo {author} {\bibfnamefont {M.}~\bibnamefont {Titze}}, \bibinfo {author} {\bibfnamefont {R.}~\bibnamefont {Haltli}}, \bibinfo {author} {\bibfnamefont {J.}~\bibnamefont {Okoro}}, \bibinfo {author} {\bibfnamefont {D.~R.}\ \bibnamefont {Tibbetts}}, \bibinfo {author} {\bibfnamefont {D.~M.}\ \bibnamefont {Udoni}}, \bibinfo {author} {\bibfnamefont {E.}~\bibnamefont {Bielejec}}, \bibinfo {author} {\bibfnamefont {M.~P.}\ \bibnamefont {Lilly}}, \bibinfo {author} {\bibfnamefont {T.-M.}\ \bibnamefont {Lu}}, \bibinfo {author} {\bibfnamefont {P.~D.~D.}\ \bibnamefont {Schwindt}},\ and\ \bibinfo {author} {\bibfnamefont {A.~M.}\ \bibnamefont {Mounce}},\ }\bibfield  {title}
  {\enquote {\bibinfo {title} {Measurement and simulation of the magnetic fields from a 555 timer integrated circuit using a quantum diamond microscope and finite-element analysis},}\ }\href@noop {} {\bibfield  {journal} {\bibinfo  {journal} {Phys. Rev. Applied}\ }\textbf {\bibinfo {volume} {17}},\ \bibinfo {pages} {014021} (\bibinfo {year} {2022})}\BibitemShut {NoStop}%
\bibitem [{\citenamefont {Arai}\ \emph {et~al.}(2022)\citenamefont {Arai}, \citenamefont {Kuwahata}, \citenamefont {Nishitani}, \citenamefont {Fujisaki}, \citenamefont {Matsuki}, \citenamefont {Nishio}, \citenamefont {Xin}, \citenamefont {Cao}, \citenamefont {Hatano}, \citenamefont {Onoda} \emph {et~al.}}]{arai2022millimetre}%
  \BibitemOpen
  \bibfield  {author} {\bibinfo {author} {\bibfnamefont {K.}~\bibnamefont {Arai}}, \bibinfo {author} {\bibfnamefont {A.}~\bibnamefont {Kuwahata}}, \bibinfo {author} {\bibfnamefont {D.}~\bibnamefont {Nishitani}}, \bibinfo {author} {\bibfnamefont {I.}~\bibnamefont {Fujisaki}}, \bibinfo {author} {\bibfnamefont {R.}~\bibnamefont {Matsuki}}, \bibinfo {author} {\bibfnamefont {Y.}~\bibnamefont {Nishio}}, \bibinfo {author} {\bibfnamefont {Z.}~\bibnamefont {Xin}}, \bibinfo {author} {\bibfnamefont {X.}~\bibnamefont {Cao}}, \bibinfo {author} {\bibfnamefont {Y.}~\bibnamefont {Hatano}}, \bibinfo {author} {\bibfnamefont {S.}~\bibnamefont {Onoda}}, \emph {et~al.},\ }\bibfield  {title} {\enquote {\bibinfo {title} {Millimetre-scale magnetocardiography of living rats with thoracotomy},}\ }\href@noop {} {\bibfield  {journal} {\bibinfo  {journal} {Communications Physics}\ }\textbf {\bibinfo {volume} {5}},\ \bibinfo {pages} {200} (\bibinfo {year} {2022})}\BibitemShut {NoStop}%
\bibitem [{\citenamefont {Fescenko}\ \emph {et~al.}(2019)\citenamefont {Fescenko}, \citenamefont {Laraoui}, \citenamefont {Smits}, \citenamefont {Mosavian}, \citenamefont {Kehayias}, \citenamefont {Seto}, \citenamefont {Bougas}, \citenamefont {Jarmola},\ and\ \citenamefont {Acosta}}]{fescenko2019diamond}%
  \BibitemOpen
  \bibfield  {author} {\bibinfo {author} {\bibfnamefont {I.}~\bibnamefont {Fescenko}}, \bibinfo {author} {\bibfnamefont {A.}~\bibnamefont {Laraoui}}, \bibinfo {author} {\bibfnamefont {J.}~\bibnamefont {Smits}}, \bibinfo {author} {\bibfnamefont {N.}~\bibnamefont {Mosavian}}, \bibinfo {author} {\bibfnamefont {P.}~\bibnamefont {Kehayias}}, \bibinfo {author} {\bibfnamefont {J.}~\bibnamefont {Seto}}, \bibinfo {author} {\bibfnamefont {L.}~\bibnamefont {Bougas}}, \bibinfo {author} {\bibfnamefont {A.}~\bibnamefont {Jarmola}},\ and\ \bibinfo {author} {\bibfnamefont {V.~M.}\ \bibnamefont {Acosta}},\ }\bibfield  {title} {\enquote {\bibinfo {title} {Diamond magnetic microscopy of malarial hemozoin nanocrystals},}\ }\href@noop {} {\bibfield  {journal} {\bibinfo  {journal} {Physical review applied}\ }\textbf {\bibinfo {volume} {11}},\ \bibinfo {pages} {034029} (\bibinfo {year} {2019})}\BibitemShut {NoStop}%
\bibitem [{\citenamefont {Horsley}\ \emph {et~al.}(2018)\citenamefont {Horsley}, \citenamefont {Appel}, \citenamefont {Wolters}, \citenamefont {Achard}, \citenamefont {Tallaire}, \citenamefont {Maletinsky},\ and\ \citenamefont {Treutlein}}]{horsley2018microwave}%
  \BibitemOpen
  \bibfield  {author} {\bibinfo {author} {\bibfnamefont {A.}~\bibnamefont {Horsley}}, \bibinfo {author} {\bibfnamefont {P.}~\bibnamefont {Appel}}, \bibinfo {author} {\bibfnamefont {J.}~\bibnamefont {Wolters}}, \bibinfo {author} {\bibfnamefont {J.}~\bibnamefont {Achard}}, \bibinfo {author} {\bibfnamefont {A.}~\bibnamefont {Tallaire}}, \bibinfo {author} {\bibfnamefont {P.}~\bibnamefont {Maletinsky}},\ and\ \bibinfo {author} {\bibfnamefont {P.}~\bibnamefont {Treutlein}},\ }\bibfield  {title} {\enquote {\bibinfo {title} {Microwave device characterization using a widefield diamond microscope},}\ }\href@noop {} {\bibfield  {journal} {\bibinfo  {journal} {Phys. Rev. Appl.}\ }\textbf {\bibinfo {volume} {10}},\ \bibinfo {pages} {044039} (\bibinfo {year} {2018})}\BibitemShut {NoStop}%
\bibitem [{\citenamefont {Wang}\ \emph {et~al.}(2015)\citenamefont {Wang}, \citenamefont {Yuan}, \citenamefont {Huang}, \citenamefont {Rong}, \citenamefont {Wang}, \citenamefont {Xu}, \citenamefont {Duan}, \citenamefont {Ju}, \citenamefont {Shi},\ and\ \citenamefont {Du}}]{wang2015high}%
  \BibitemOpen
  \bibfield  {author} {\bibinfo {author} {\bibfnamefont {P.}~\bibnamefont {Wang}}, \bibinfo {author} {\bibfnamefont {Z.}~\bibnamefont {Yuan}}, \bibinfo {author} {\bibfnamefont {P.}~\bibnamefont {Huang}}, \bibinfo {author} {\bibfnamefont {X.}~\bibnamefont {Rong}}, \bibinfo {author} {\bibfnamefont {M.}~\bibnamefont {Wang}}, \bibinfo {author} {\bibfnamefont {X.}~\bibnamefont {Xu}}, \bibinfo {author} {\bibfnamefont {C.}~\bibnamefont {Duan}}, \bibinfo {author} {\bibfnamefont {C.}~\bibnamefont {Ju}}, \bibinfo {author} {\bibfnamefont {F.}~\bibnamefont {Shi}},\ and\ \bibinfo {author} {\bibfnamefont {J.}~\bibnamefont {Du}},\ }\bibfield  {title} {\enquote {\bibinfo {title} {High-resolution vector microwave magnetometry based on solid-state spins in diamond},}\ }\href@noop {} {\bibfield  {journal} {\bibinfo  {journal} {Nature communications}\ }\textbf {\bibinfo {volume} {6}},\ \bibinfo {pages} {6631} (\bibinfo {year} {2015})}\BibitemShut {NoStop}%
\bibitem [{\citenamefont {Appel}\ \emph {et~al.}(2015)\citenamefont {Appel}, \citenamefont {Ganzhorn}, \citenamefont {Neu},\ and\ \citenamefont {Maletinsky}}]{appel2015nanoscale}%
  \BibitemOpen
  \bibfield  {author} {\bibinfo {author} {\bibfnamefont {P.}~\bibnamefont {Appel}}, \bibinfo {author} {\bibfnamefont {M.}~\bibnamefont {Ganzhorn}}, \bibinfo {author} {\bibfnamefont {E.}~\bibnamefont {Neu}},\ and\ \bibinfo {author} {\bibfnamefont {P.}~\bibnamefont {Maletinsky}},\ }\bibfield  {title} {\enquote {\bibinfo {title} {Nanoscale microwave imaging with a single electron spin in diamond},}\ }\href@noop {} {\bibfield  {journal} {\bibinfo  {journal} {New J. Phys.}\ }\textbf {\bibinfo {volume} {17}},\ \bibinfo {pages} {112001} (\bibinfo {year} {2015})}\BibitemShut {NoStop}%
\bibitem [{sup()}]{suppl}%
  \BibitemOpen
  \href@noop {} {}\bibinfo {note} {Additional details are included in the supplemental material.}\BibitemShut {Stop}%
\bibitem [{\citenamefont {Magaletti}\ \emph {et~al.}(2024)\citenamefont {Magaletti}, \citenamefont {Mayer}, \citenamefont {Roch},\ and\ \citenamefont {Debuisschert}}]{magaletti2024modelling}%
  \BibitemOpen
  \bibfield  {author} {\bibinfo {author} {\bibfnamefont {S.}~\bibnamefont {Magaletti}}, \bibinfo {author} {\bibfnamefont {L.}~\bibnamefont {Mayer}}, \bibinfo {author} {\bibfnamefont {J.-F.}\ \bibnamefont {Roch}},\ and\ \bibinfo {author} {\bibfnamefont {T.}~\bibnamefont {Debuisschert}},\ }\bibfield  {title} {\enquote {\bibinfo {title} {Modelling rabi oscillations for widefield radiofrequency imaging in nitrogen-vacancy centers in diamond},}\ }\href@noop {} {\bibfield  {journal} {\bibinfo  {journal} {New J. Phys.}\ }\textbf {\bibinfo {volume} {26}},\ \bibinfo {pages} {023020} (\bibinfo {year} {2024})}\BibitemShut {NoStop}%
\bibitem [{\citenamefont {Taylor}\ \emph {et~al.}(2008)\citenamefont {Taylor}, \citenamefont {Cappellaro}, \citenamefont {Childress}, \citenamefont {Jiang}, \citenamefont {Budker}, \citenamefont {Hemmer}, \citenamefont {Yacoby}, \citenamefont {Walsworth},\ and\ \citenamefont {Lukin}}]{dima_magReview}%
  \BibitemOpen
  \bibfield  {author} {\bibinfo {author} {\bibfnamefont {J.~M.}\ \bibnamefont {Taylor}}, \bibinfo {author} {\bibfnamefont {P.}~\bibnamefont {Cappellaro}}, \bibinfo {author} {\bibfnamefont {L.}~\bibnamefont {Childress}}, \bibinfo {author} {\bibfnamefont {L.}~\bibnamefont {Jiang}}, \bibinfo {author} {\bibfnamefont {D.}~\bibnamefont {Budker}}, \bibinfo {author} {\bibfnamefont {P.~R.}\ \bibnamefont {Hemmer}}, \bibinfo {author} {\bibfnamefont {A.}~\bibnamefont {Yacoby}}, \bibinfo {author} {\bibfnamefont {R.}~\bibnamefont {Walsworth}},\ and\ \bibinfo {author} {\bibfnamefont {M.~D.}\ \bibnamefont {Lukin}},\ }\bibfield  {title} {\enquote {\bibinfo {title} {High-sensitivity diamond magnetometer with nanoscale resolution},}\ }\href@noop {} {\bibfield  {journal} {\bibinfo  {journal} {Nat. Phys.}\ }\textbf {\bibinfo {volume} {4}},\ \bibinfo {pages} {810--816} (\bibinfo {year} {2008})}\BibitemShut {NoStop}%
\bibitem [{\citenamefont {Rondin}\ \emph {et~al.}(2014)\citenamefont {Rondin}, \citenamefont {Tetienne}, \citenamefont {Hingan}, \citenamefont {Roch}, \citenamefont {Maletinsky},\ and\ \citenamefont {Jacques}}]{Rondin}%
  \BibitemOpen
  \bibfield  {author} {\bibinfo {author} {\bibfnamefont {L.}~\bibnamefont {Rondin}}, \bibinfo {author} {\bibfnamefont {J.-P.}\ \bibnamefont {Tetienne}}, \bibinfo {author} {\bibfnamefont {T.}~\bibnamefont {Hingan}}, \bibinfo {author} {\bibfnamefont {J.-F.}\ \bibnamefont {Roch}}, \bibinfo {author} {\bibfnamefont {P.}~\bibnamefont {Maletinsky}},\ and\ \bibinfo {author} {\bibfnamefont {V.}~\bibnamefont {Jacques}},\ }\bibfield  {title} {\enquote {\bibinfo {title} {Magnetometry with nitrogen-vacancy defects in diamond},}\ }\href@noop {} {\bibfield  {journal} {\bibinfo  {journal} {Rep. Prog. Phys.}\ }\textbf {\bibinfo {volume} {77}},\ \bibinfo {pages} {056503} (\bibinfo {year} {2014})}\BibitemShut {NoStop}%
\bibitem [{\citenamefont {Chen}\ \emph {et~al.}(2020)\citenamefont {Chen}, \citenamefont {Hou}, \citenamefont {Ge}, \citenamefont {Zhang}, \citenamefont {Ji}, \citenamefont {Li}, \citenamefont {Qian}, \citenamefont {Wang}, \citenamefont {Xu},\ and\ \citenamefont {Du}}]{chen2020calibration}%
  \BibitemOpen
  \bibfield  {author} {\bibinfo {author} {\bibfnamefont {B.}~\bibnamefont {Chen}}, \bibinfo {author} {\bibfnamefont {X.}~\bibnamefont {Hou}}, \bibinfo {author} {\bibfnamefont {F.}~\bibnamefont {Ge}}, \bibinfo {author} {\bibfnamefont {X.}~\bibnamefont {Zhang}}, \bibinfo {author} {\bibfnamefont {Y.}~\bibnamefont {Ji}}, \bibinfo {author} {\bibfnamefont {H.}~\bibnamefont {Li}}, \bibinfo {author} {\bibfnamefont {P.}~\bibnamefont {Qian}}, \bibinfo {author} {\bibfnamefont {Y.}~\bibnamefont {Wang}}, \bibinfo {author} {\bibfnamefont {N.}~\bibnamefont {Xu}},\ and\ \bibinfo {author} {\bibfnamefont {J.}~\bibnamefont {Du}},\ }\bibfield  {title} {\enquote {\bibinfo {title} {Calibration-free vector magnetometry using nitrogen-vacancy center in diamond integrated with optical vortex beam},}\ }\href@noop {} {\bibfield  {journal} {\bibinfo  {journal} {Nano Letters}\ }\textbf {\bibinfo {volume} {20}},\ \bibinfo {pages} {8267--8272} (\bibinfo {year} {2020})}\BibitemShut {NoStop}%
\bibitem [{\citenamefont {Kazi}\ \emph {et~al.}(2021)\citenamefont {Kazi}, \citenamefont {Shelby}, \citenamefont {Watanabe}, \citenamefont {Itoh}, \citenamefont {Shutthanandan}, \citenamefont {Wiggins},\ and\ \citenamefont {Fu}}]{kazi2021wide}%
  \BibitemOpen
  \bibfield  {author} {\bibinfo {author} {\bibfnamefont {Z.}~\bibnamefont {Kazi}}, \bibinfo {author} {\bibfnamefont {I.~M.}\ \bibnamefont {Shelby}}, \bibinfo {author} {\bibfnamefont {H.}~\bibnamefont {Watanabe}}, \bibinfo {author} {\bibfnamefont {K.~M.}\ \bibnamefont {Itoh}}, \bibinfo {author} {\bibfnamefont {V.}~\bibnamefont {Shutthanandan}}, \bibinfo {author} {\bibfnamefont {P.~A.}\ \bibnamefont {Wiggins}},\ and\ \bibinfo {author} {\bibfnamefont {K.-M.~C.}\ \bibnamefont {Fu}},\ }\bibfield  {title} {\enquote {\bibinfo {title} {Wide-field dynamic magnetic microscopy using double-double quantum driving of a diamond defect ensemble},}\ }\href@noop {} {\bibfield  {journal} {\bibinfo  {journal} {Phys. Rev. Appl.}\ }\textbf {\bibinfo {volume} {15}},\ \bibinfo {pages} {054032} (\bibinfo {year} {2021})}\BibitemShut {NoStop}%
\bibitem [{\citenamefont {Hart}\ \emph {et~al.}(2021)\citenamefont {Hart}, \citenamefont {Schloss}, \citenamefont {Turner}, \citenamefont {Scheidegger}, \citenamefont {Bauch},\ and\ \citenamefont {Walsworth}}]{hart2021n}%
  \BibitemOpen
  \bibfield  {author} {\bibinfo {author} {\bibfnamefont {C.~A.}\ \bibnamefont {Hart}}, \bibinfo {author} {\bibfnamefont {J.~M.}\ \bibnamefont {Schloss}}, \bibinfo {author} {\bibfnamefont {M.~J.}\ \bibnamefont {Turner}}, \bibinfo {author} {\bibfnamefont {P.~J.}\ \bibnamefont {Scheidegger}}, \bibinfo {author} {\bibfnamefont {E.}~\bibnamefont {Bauch}},\ and\ \bibinfo {author} {\bibfnamefont {R.~L.}\ \bibnamefont {Walsworth}},\ }\bibfield  {title} {\enquote {\bibinfo {title} {N-v--diamond magnetic microscopy using a double quantum 4-ramsey protocol},}\ }\href@noop {} {\bibfield  {journal} {\bibinfo  {journal} {Phys. Rev. Appl.}\ }\textbf {\bibinfo {volume} {15}},\ \bibinfo {pages} {044020} (\bibinfo {year} {2021})}\BibitemShut {NoStop}%
\end{thebibliography}%


%aipnum4-2.bst 2019-01-14 (MD) hand-edited version of apsrev4-1.bst
\begin{thebibliography}{4}%
\makeatletter
\providecommand \@ifxundefined [1]{%
 \@ifx{#1\undefined}
}%
\providecommand \@ifnum [1]{%
 \ifnum #1\expandafter \@firstoftwo
 \else \expandafter \@secondoftwo
 \fi
}%
\providecommand \@ifx [1]{%
 \ifx #1\expandafter \@firstoftwo
 \else \expandafter \@secondoftwo
 \fi
}%
\providecommand \natexlab [1]{#1}%
\providecommand \enquote  [1]{``#1''}%
\providecommand \bibnamefont  [1]{#1}%
\providecommand \bibfnamefont [1]{#1}%
\providecommand \citenamefont [1]{#1}%
\providecommand \href@noop [0]{\@secondoftwo}%
\providecommand \href [0]{\begingroup \@sanitize@url \@href}%
\providecommand \@href[1]{\@@startlink{#1}\@@href}%
\providecommand \@@href[1]{\endgroup#1\@@endlink}%
\providecommand \@sanitize@url [0]{\catcode `\\12\catcode `\$12\catcode `\&12\catcode `\#12\catcode `\^12\catcode `\_12\catcode `\%12\relax}%
\providecommand \@@startlink[1]{}%
\providecommand \@@endlink[0]{}%
\providecommand \url  [0]{\begingroup\@sanitize@url \@url }%
\providecommand \@url [1]{\endgroup\@href {#1}{\urlprefix }}%
\providecommand \urlprefix  [0]{URL }%
\providecommand \Eprint [0]{\href }%
\providecommand \doibase [0]{https://doi.org/}%
\providecommand \selectlanguage [0]{\@gobble}%
\providecommand \bibinfo  [0]{\@secondoftwo}%
\providecommand \bibfield  [0]{\@secondoftwo}%
\providecommand \translation [1]{[#1]}%
\providecommand \BibitemOpen [0]{}%
\providecommand \bibitemStop [0]{}%
\providecommand \bibitemNoStop [0]{.\EOS\space}%
\providecommand \EOS [0]{\spacefactor3000\relax}%
\providecommand \BibitemShut  [1]{\csname bibitem#1\endcsname}%
\let\auto@bib@innerbib\@empty
%</preamble>
\bibitem [{\citenamefont {Horsley}\ \emph {et~al.}(2018)\citenamefont {Horsley}, \citenamefont {Appel}, \citenamefont {Wolters}, \citenamefont {Achard}, \citenamefont {Tallaire}, \citenamefont {Maletinsky},\ and\ \citenamefont {Treutlein}}]{horsley2018microwave}%
  \BibitemOpen
  \bibfield  {author} {\bibinfo {author} {\bibfnamefont {A.}~\bibnamefont {Horsley}}, \bibinfo {author} {\bibfnamefont {P.}~\bibnamefont {Appel}}, \bibinfo {author} {\bibfnamefont {J.}~\bibnamefont {Wolters}}, \bibinfo {author} {\bibfnamefont {J.}~\bibnamefont {Achard}}, \bibinfo {author} {\bibfnamefont {A.}~\bibnamefont {Tallaire}}, \bibinfo {author} {\bibfnamefont {P.}~\bibnamefont {Maletinsky}},\ and\ \bibinfo {author} {\bibfnamefont {P.}~\bibnamefont {Treutlein}},\ }\bibfield  {title} {\enquote {\bibinfo {title} {Microwave device characterization using a widefield diamond microscope},}\ }\href@noop {} {\bibfield  {journal} {\bibinfo  {journal} {Phys. Rev. Appl.}\ }\textbf {\bibinfo {volume} {10}},\ \bibinfo {pages} {044039} (\bibinfo {year} {2018})}\BibitemShut {NoStop}%
\bibitem [{\citenamefont {Appel}\ \emph {et~al.}(2015)\citenamefont {Appel}, \citenamefont {Ganzhorn}, \citenamefont {Neu},\ and\ \citenamefont {Maletinsky}}]{appel2015nanoscale}%
  \BibitemOpen
  \bibfield  {author} {\bibinfo {author} {\bibfnamefont {P.}~\bibnamefont {Appel}}, \bibinfo {author} {\bibfnamefont {M.}~\bibnamefont {Ganzhorn}}, \bibinfo {author} {\bibfnamefont {E.}~\bibnamefont {Neu}},\ and\ \bibinfo {author} {\bibfnamefont {P.}~\bibnamefont {Maletinsky}},\ }\bibfield  {title} {\enquote {\bibinfo {title} {Nanoscale microwave imaging with a single electron spin in diamond},}\ }\href@noop {} {\bibfield  {journal} {\bibinfo  {journal} {New J. Phys.}\ }\textbf {\bibinfo {volume} {17}},\ \bibinfo {pages} {112001} (\bibinfo {year} {2015})}\BibitemShut {NoStop}%
\bibitem [{\citenamefont {Yamaguchi}\ \emph {et~al.}(2019)\citenamefont {Yamaguchi}, \citenamefont {Matsuzaki}, \citenamefont {Saito}, \citenamefont {Saijo}, \citenamefont {Watanabe}, \citenamefont {Mizuochi},\ and\ \citenamefont {Ishi-Hayase}}]{yamaguchi2019bandwidth}%
  \BibitemOpen
  \bibfield  {author} {\bibinfo {author} {\bibfnamefont {T.}~\bibnamefont {Yamaguchi}}, \bibinfo {author} {\bibfnamefont {Y.}~\bibnamefont {Matsuzaki}}, \bibinfo {author} {\bibfnamefont {S.}~\bibnamefont {Saito}}, \bibinfo {author} {\bibfnamefont {S.}~\bibnamefont {Saijo}}, \bibinfo {author} {\bibfnamefont {H.}~\bibnamefont {Watanabe}}, \bibinfo {author} {\bibfnamefont {N.}~\bibnamefont {Mizuochi}},\ and\ \bibinfo {author} {\bibfnamefont {J.}~\bibnamefont {Ishi-Hayase}},\ }\bibfield  {title} {\enquote {\bibinfo {title} {Bandwidth analysis of ac magnetic field sensing based on electronic spin double-resonance of nitrogen-vacancy centers in diamond},}\ }\href@noop {} {\bibfield  {journal} {\bibinfo  {journal} {Jpn. J. Appl. Phys.}\ }\textbf {\bibinfo {volume} {58}},\ \bibinfo {pages} {100901} (\bibinfo {year} {2019})}\BibitemShut {NoStop}%
\bibitem [{\citenamefont {Cohen-Tannoudji}, \citenamefont {Diu},\ and\ \citenamefont {Laloë}(2019)}]{Cohen}%
  \BibitemOpen
  \bibfield  {author} {\bibinfo {author} {\bibfnamefont {C.}~\bibnamefont {Cohen-Tannoudji}}, \bibinfo {author} {\bibfnamefont {B.}~\bibnamefont {Diu}},\ and\ \bibinfo {author} {\bibfnamefont {F.}~\bibnamefont {Laloë}},\ }\href@noop {} {\emph {\bibinfo {title} {Quantum Mechanics, Volume I}}},\ \bibinfo {edition} {2nd}\ ed.\ (\bibinfo  {publisher} {Hermann and John Wiley and Sons},\ \bibinfo {year} {2019})\BibitemShut {NoStop}%
\end{thebibliography}%

% \subfile{Supplemental}

\clearpage
% \centering
% \includegraphics[scale=1.0,page=1]{Supplemental.pdf}

% \begin{figure}[t] 
% \centering
% \includegraphics[scale=0.92]{Supplemental.pdf}
% \end{figure} 

% \includepdf[pages=1]{Supplemental.pdf} 
% \clearpage
% \includepdf[pages=2]{Supplemental.pdf} 
% \clearpage
% \includepdf[pages=3]{Supplemental.pdf} 
% \clearpage
% \includepdf[pages=4]{Supplemental.pdf} 
% \clearpage
% \includepdf[pages=5]{Supplemental.pdf} 
% \clearpage
% \includepdf[pages=6]{Supplemental.pdf} 

\end{document}

\title{Supplementary Material: Wide-field microwave magnetic field imaging with nitrogen-vacancy centers in diamond}

\date{\today}
\author{Luca Basso}\email{lbasso@sandia.gov}\affiliation{\cint}
\author{Pauli Kehayias}\affiliation{\snl}\affiliation{\llMIT}
\author{Jacob Henshaw}\affiliation{\cint}\affiliation{\snl}
\author{Gajadhar Joshi}\affiliation{\cint}
\author{Michael P. Lilly}\affiliation{\cint}
\author{Matthew B. Jordan}\affiliation{\snl}
\author{Andrew M. Mounce}\email{ammounce@sandia.gov}\affiliation{\cint}
%TC:ignore

\maketitle

% \begin{figure*}[ht]
% \centering
% \includegraphics[width=0.9\linewidth]{Figure1.pdf}
% \captionsetup{width=1\linewidth}
% \caption{(a) Schematics of the AC wide-field magnetic field imaging experimental setup. A 532 nm laser is used to excite the NV centers, while a copper omega-loop delivers the MW fields to manipulate the spin states. The emitted PL is long-passed  filtered before being collected by the camera for wide field imaging. A bias magnetic field $B_{DC}$ is applied along the [111] oriented NV axis ($\sim$ 55 $^{\circ}$ from the vertical) by a pair of permanent magnets. (b) Detail of the diamond-DUT integration method. The diamond is glued on the omega-loop which is in turn mounted on a 5-axis ($x, y, z,\theta, \phi$) stage, used to bring the diamond in proximity of the DUT. (c) Optical picture of the omega-loop with the diamond glued on top. (d) Optical picture of the DUT, consisting of a 100$\upmu$m-wide copper wire. (e) NV center ground-state energy level, showing the effect of a static field $B_{DC}$ lifting the $\ket{\pm 1}$ degeneracy and the $\sigma_{\pm}$ polarized MW transitions. (f) Experimental pulse sequence for wide-field Rabi oscillations measurement.}
% \label{Figure1}
% \end{figure*}

\section{Experimental details}
The 532 nm laser excitation is provided by a solid-state diode laser (Sprout-G-5W). An acousto-optic modulator (AOM) is used to pulse the laser. The laser beam is focused into the AOM with a $f = 500$ mm lens and then collimated by another $f = 500$ mm lens. An iris is used to block the zero-order diffracted beam and allow only the first-order one to go trough. 

The AOM is driven by radio frequency (RF) pulses generated by a TPI-1002-A signal generator, gated by a switch (ZASWA-2-50DRA+) and finally amplified with a ZHL-03-5WF+ amplifier. The AOM RF drive and the switch are triggered using transistortransistor logic (TTL) pulses from a PulseBlaster ESR-Pro 500. 

To rotate the polarization of the laser beam, a half-wave ($\lambda$/2) plate is used after the iris. After the wave-plate, the beam is firstly expanded then focused with a $f = 200$ mm achromatic lens into the back aperture of a 20X/0.4NA objective to achieve a uniform illumination in a $\sim 340\times340 \upmu$m$^2$ area on the diamond surface. The laser power on the diamond sample is $\sim 2.2 W$.

The NV fluorescence is collected by the objective, filtered by a 650 nm long pass filter and finally collected by a CMOS camera equipped with $f = 200$ mm tube lens to keep the effective magnification of the objective at 20X. During the alignment of the bias magnetic field along a single NV axis, which is done through optically detected magnetic resonance (ODMR) spectroscopy, the camera is substituted with a photodiode and the signal is processed with a lock-in amplifier.

The microwave (MW) pulses to control the NVs spin states are delivered to the NV layer by a fabricated MW-loop made by a 16$\upmu$m-thick Cu layer on a 0.8mm-thick FR4 substrate on which the diamond is glued on. The MW-loop is fabricated starting from a double-sided Cu-clad FR4 board that we cut into the MW-loop shape with a laser-cutting machine. The DUT is fabricated trough a standard photolitography process: a Al$_2$O$_3$ substrate is spin coated with photoresit (AZ5
14), exposed to form a 100$\upmu$m-wide stripline pattern, then developed (MF319). The substrate is then deposited with a 3nm-thick layer of Ti (for improved adhesion) and 300nm-thick Cu layer. Finally, the substrate is rinsed in acetone to remove the photoresist and leave the Cu stripline.

For both the MW-loop and the DUT, the MW pulses are generated by two different TPI-1002-A signal generators, gated using switches and TTL pulses, and separately amplified .

The DUT is mounted on a xyz linear stage whereas the diamond/MW-loop is mounted on another xyz stage on which a rotation stage and a goniometer are mounted. In this way we can precisely control the position as well as two rotation angles (namely tip and tilt) of the diamond/MW-loop for a better positioning in proximity of the DUT.

The bias magnetic field is provided by two Nd permanent magnets each mounted on a motorized linear stage used to precisely control the relative distance of the two magnets, i.e to control the amplitude of the bias field. The two linear stages are mounted on the same motorized rotation stage which is in turn mounted on another precision rotation stage. With the serial combination of the two rotation stages we have an automatic control the two rotation axis, namely the polar angle $\theta$ and the azimuthal angle $\phi$.

The NV diamond sensor is a 4-$\upmu$m-thick, $^{12}$C-enriched diamond layer, doped with 25 ppm of $^{14}$N overgrown on a 4×4×0.5 mm$^3$ electronic grade diamond (Element 6, native nitrogen density < 5 ppb).
For vacancies formation the overgrwon diamond is irradiated with a 1 MeV, 1.2e18 cm$^{-2}$ ebeam. For NV activation the diamond is annealed in a ultra-high vacuum furnace ($<$ 10$^{-8}$ Torr) with the following recipe: 2h ramp to 400$^{\circ}$C, 2h soak, 2h ramp to 550$^{\circ}$C, 2h soak, 2h ramp to 800$^{\circ}$C, 4h soak, 2h ramp to 1100$^{\circ}$C, 2h soak, and cooldown to room temperature. Finally, the diamond is cleaned with triacid solution (1:1:1 sulfuric, nitric and perchloric acids) at 250 $^{\circ}$C for 1h.

\section{Fitting of Rabi oscillation data}
\begin{figure*}[ht]
\centering
\includegraphics[width=0.9\linewidth]{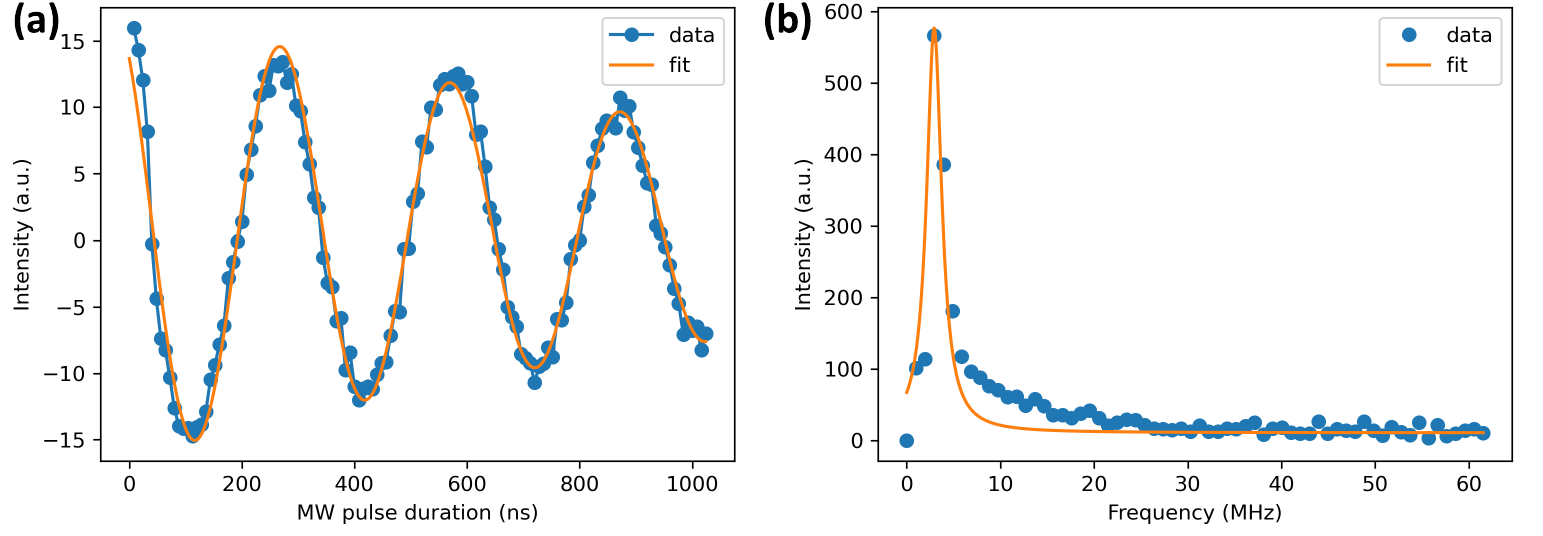}
\captionsetup{width=1\linewidth}
\caption{(a) Rabi oscillations experimental data (blue dots) and fit trough Eq. (2) of main text (orange solid lines). (b) FFT of the experimental data (blue dots) and its fit with a Lorentzial function (orange solid line). The fitting parameters of the FFT vector (b) are used as guess parameters for the fit of the Rabi oscillation shown in (a).}
\label{FigureS1}
\end{figure*}

To fit the experimental data (blue dots in Fig. \ref{FigureS1}(a)), we firstly calculate its fast Fourier transform (FFT) (blue dots in Fig. \ref{FigureS1}(b)). Then, we fit the Fourier-transformed Rabi oscillation with a Lorentzian function (solid orange line in Fig. \ref{FigureS1}(b)). We use the center and the width of the Lorentzian curve as guess parameters for, respectively, the fitting parameters $\Omega_+$ and $1/\tau_R$ of Eq. (2) of main text. Finally, we fit the Rabi oscillation with Eq. (2) of main text (solid orange line in Fig. \ref{FigureS1}(a)).

\section{Projection of the MW magnetic field on the NV axis}
In the following we describe, as similarly done by\cite[S][]{ horsley2018microwave}, how the MW magnetic field generated by the DUT in the laboratory frame $\left\{ x,y,z \right \}$ projects in the NV reference frame $\left\{ x',y',z' \right \}$, defined with $\widehat{e}_{z'}$ directed along the [111] NV axis. At the position of the NV spins $\vec{r} = (x,y,z)$ in the laboratory reference frame, the DUT generates an MW magnetic field $\vec{B}_{MW}(\vec{r}) \cos (\omega_{MW} t)$ that can be decomposed as
\begin{equation}
   \vec{B}_{MW}(\vec{r}) = \begin{pmatrix} B_{x,MW}(\vec{r}) \\ B_{y,MW}(\vec{r}) \\  B_{z,MW}(\vec{r}) \end{pmatrix}
\label{B_expan}
\end{equation}
where the frequency component $\cos (\omega_{MW} t)$ has been neglected for simplicity. The projection of $\vec{B}_{MW}(\vec{r})$ on the NV reference frame, defined as $\vec{B'}_{MW}(\vec{r'})$, is given by:
\begin{equation}
   \begin{pmatrix} B'_{x',MW}(\vec{r'}) \\ B'_{y',MW}(\vec{r'}) \\  B'_{z',MW}(\vec{r'}) \end{pmatrix} = R(\theta, \phi) \begin{pmatrix} B_{x,MW}(\vec{r}) \\ B_{y,MW}(\vec{r}) \\  B_{z,MW}(\vec{r}) \end{pmatrix}
\label{B_proj}
\end{equation}
where $R(\theta, \phi)$ is the transformation matrix from the laboratory to the NV reference frame such that $\vec{r'}=R(\theta, \phi)\vec{r}$, and has the following form:
\begin{equation}
   R(\theta,\phi) = \begin{pmatrix}
   \cos \theta \cos \phi  &  \cos \theta \sin \phi  & - \sin \theta \\
   -  \sin \phi  & \cos \phi  & 0 \\
    \sin \theta  \cos \phi &  \sin \theta \sin \phi &  \cos \theta
\end{pmatrix}
\label{R_matrix}
\end{equation}
The angles $\theta$ and $\phi$ are respectively the polar and azimuthal angles, as defined in Fig. 1 of main text. The [111] NV axis is the axis along which the bias $B_0$ field is aligned, and in our particular case is oriented respect to the laboratory reference frame with $\theta = 55^{\circ}$ and $\phi = 87^{\circ}$. Putting together Eq. \ref{B_proj} and \ref{R_matrix} we obtain:
\begin{equation}
   \begin{pmatrix} B'_{x',MW}(\vec{r'}) \\ B'_{y',MW}(\vec{r'}) \\  B'_{z',MW}(\vec{r'}) \end{pmatrix} = \begin{pmatrix}
   B_{x,MW}(\vec{r}) \cos \theta \cos \phi  + B_{y,MW}(\vec{r}) \cos \theta \sin \phi   -  B_{z,MW}(\vec{r})\sin \theta \\
   - B_{x,MW}(\vec{r}) \sin \phi + B_{y,MW}(\vec{r})   \cos \phi   \\
   B_{x,MW}(\vec{r})  \sin \theta  \cos \phi +B_{y,MW}(\vec{r})   \sin \theta \sin \phi  +B_{z,MW}(\vec{r})  \cos \theta
\end{pmatrix}
\label{B_NVref}
\end{equation}
The NVs spin transitions are only driven by the components of $\vec{B'}_{MW}$ orthogonal to the NV axis, that we define as $\vec{B'}_{\perp ,MW}$. This component is linearly polarized thus it can be written as a linear combination of left-handed and right-handed circularly polarized fields:
\begin{equation}
   B'_{\perp ,MW}(\vec{r'}) = \frac {1} {\sqrt{2}} B'_{- ,MW}(\vec{r'}) + \frac {i} {\sqrt{2}} B'_{+ ,MW}(\vec{r'})
\label{B_lin_comb}
\end{equation}
where $\vec{B'}_{+,MW}$ and $\vec{B'}_{-,MW}$ are respectively the left-handed and right-handed magnetic field components of the transitions $\sigma_+$ and $\sigma_-$.
Finally, the left- (right)-handed components can be written in the NV reference frame as:
\begin{equation}
   B'_{\mp ,MW}(\vec{r'}) = \vert B'_{x',MW}(\vec{r'}) \pm i B'_{y' ,MW}(\vec{r'}) \vert
\label{B_+-}
\end{equation}

\section{MW field generated by a stripline}
In this section we give the analytical expression, adapted from \cite{appel2015nanoscale}, for the MW field generated by a stripline used to fit the measured $B_{-,MW}$ as reported in Fig. 2(f) of main text. We assume that the current density in the stripline is homogeneous and flow in only one direction, namely $\Vec{J} = (J_x = J,0,0)$, and that the thickness of stripline $t$ is much smaller than the stripline width $w$, namely the stripline is infinitely thin. If we choose the origin of laboratory reference frame to be in the center of the stripline, with the $z$ axis normal to the stripline surface and $x$ parallel to the stripline, the magnetic vector potential $A_{x}(\vec{r}, J)$ oriented along $x$ on a NV at a position $\vec{r} = (y, z = d + t/2)$ is:
\begin{equation}
    A_{x}(\vec{r}, J) = \frac{\mu_0}{4\pi} \int \frac{J(y',z')}{\vert \vec{r}-\vec{r'} \vert} d^3\vec{r'}
\label{vec_pot_0}
\end{equation}
where $d$ is the stand-off distance between the NV layer and the stripline while $\vec{r'} = (x',y',z')$ is the coordinate of the MW current density inside the stripline. This equation can be solved in analogy to an electrostatic potential $\Phi$ with a charge distribution $\rho(y',z')$:
\begin{equation}
   \Phi = \frac{1}{4\pi\epsilon_0} \int \frac{\rho(y',z')}{\vert \vec{r}-\vec{r'} \vert} d^3\vec{r'} = -\frac{1}{2\pi\epsilon_0} \iint \rho(y',z') \ln \left( \frac{\sqrt{(y-y')^2+(z-z')^2}}{r_0} \right) d{y'}d{z'}
\label{elec_pot_charge_dist}
\end{equation}
where $r_0$ is a integration constant. Thus, Eq.\ref{vec_pot_0} can be rewritten as:
\begin{equation}
    A_{x}(\vec{r}, J) =  -\frac{\mu_0}{2\pi} \iint J(y',z') \ln \left( \frac{\sqrt{(y-y')^2+(z-z')^2}}{r_0} \right) d{y'}d{z'}
\label{vec_pot_1}
\end{equation}
Taking into account the assumption that the stripline is infinitely thin, the integration on $z'$ only results $t$, whereas the assumption that the current density is uniform allows to write $J(y',z') = J$. This leads to:
\begin{equation}
    A_{x}(\vec{r}, J) =  -\frac{\mu_0 J t}{2\pi} \int_{-w/2}^{w/2} \ln \left( \sqrt{(y-y')^2+z^2} \right) d{y'}
\label{vec_pot_2}
\end{equation}
The magnetic field resulting from the above magnetic vector potential can be found through:
\begin{equation}
   \vec{B}_{MW}(\vec{r}, J) = - \widehat{e}_x \times \nabla A_{x}(\vec{r}, J) = \begin{pmatrix}  0 \\ -\partial_z A_{x}(\vec{r}, J) \\  \partial_y A_{x}(\vec{r}, J) \end{pmatrix} = \begin{pmatrix} B_x \\ B_y \\  B_z \end{pmatrix}
\label{B_from_vec_pot_2}
\end{equation}
The above magnetic field in the laboratory reference frame can be projected in the NV reference frame trough Eq. \ref{B_NVref}, leading to:
\begin{equation}
   \vec{B'}_{MW}(\vec{r}, J) = \begin{pmatrix}  -\partial_z A_{x}(\vec{r}, J) \sin \phi \cos \theta -\partial_y A_{x}(\vec{r}, J) \sin \theta \\  \partial_z A_{x}(\vec{r}, J) \cos \phi \\  -\partial_z A_{x}(\vec{r}, J) \sin \phi \sin \theta +\partial_y A_{x}(\vec{r}, J) \cos \theta\end{pmatrix} 
\label{B_matrix}
\end{equation}
Finally, we can put together Eq. \ref{B_+-} with \ref{B_matrix} to obtain:
\begin{equation}
    \begin{aligned}
   \vec{B'}_{-,MW}(\vec{r}, J, \theta , \phi) & = \vert \vec{B'}_{x,MW}(\vec{r}, J, \theta , \phi) + i \vec{B'}_{y,MW}(\vec{r}, J, \theta , \phi) \vert 
   \\
  & = \vert -\partial_z A_{x}(\vec{r}, J) \sin \phi \cos \theta -\partial_y A_{x}(\vec{r}, J) \sin \theta + i  \partial_z A_{x}(\vec{r}, J) \cos \phi \vert
\label{B_fit_equation}
    \end{aligned}
\end{equation}
The above equation is used as the fitting function with $J$ and $d$ as fitting parameters, whereas $\theta$ and $\phi$ are the angles over which the bias magnetic field is oriented to be aligned to the [111] NV axis.

\section{Rabi oscillations from two MW magnetic field sources}
In this section we will discuss Rabi oscillations in the presence of two MW sources (i.e. the MW loop and the the DUT) and the derivation of Eq. 2 of the main text. The two MW sources generate a total magnetic field oscillating with frequency $\omega$ in the laboratory reference frame $\vec{B}_{MW}(t) = \vec{B}_{LOOP}(t) + \vec{B}_{DUT}(t)$, where both $\vec{B}_{LOOP}(t)$ and $\vec{B}_{DUT}(t)$ has the form of Eq. \ref{B_expan}. With the bias field $B_0$ aligned to one of the NV axis, direction we define as $z_{NV}$, the NV ground-state Hamiltonian can be written as \cite{yamaguchi2019bandwidth}: 
\begin{equation}
   H_{NV} / h =H_0 + H_{MW}(t) \approx D_{GS} S_z^2 + \gamma_{NV} B_0 S_z  + \gamma_{NV}  \bold{B_{MW}(t)} \cdot \bold{S}
\label{Eq_hamiltonian}
\end{equation}
\begin{figure*}[b]
\centering
\includegraphics[width=0.5\linewidth]{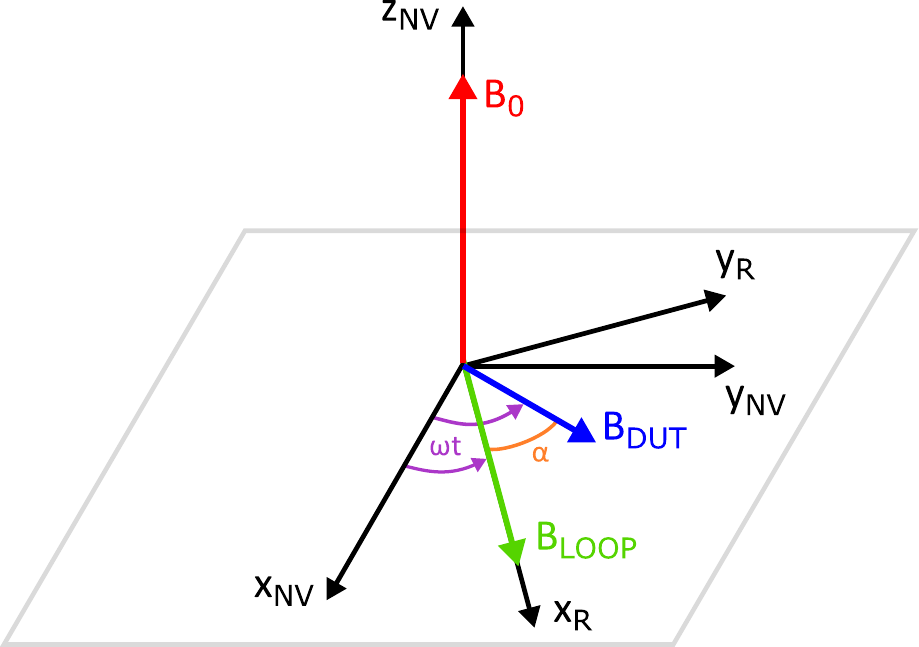}
\captionsetup{width=1\linewidth}
\caption{NV reference frame $\left \{x_{NV},y_{NV},z_{NV}\right \}$ and frame $\left \{x_{R},y_{R},z_{NV}\right \}$ rotating with an angular frequency $\omega$ with rotation axis parallel to $z_{NV}$. The direction of $z_{NV}$ is set by the direction of the bias field $B_0$ (red arrow), which is turn oriented along the NV [111] crystal orientation.  The green and blue arrows are respectively the perpendicular component of $\vec{B}_{LOOP}(t)$ and $\vec{B}_{DUT}(t)$.}
\label{FigureS2}
\end{figure*}
where $h$ is the Planck constant, $\bold{S} = (S_x, S_y, S_z)$ is the vector of spin-1 Pauli matrices, and we neglected other terms in the Hamiltonian, such as hyperfine interaction with nearby nuclear spins or presence of strain in the diamond structure. We notice that only the component of $\vec{B}_{MW}(t)$ perpendicular to the quantization axis $z_{NV}$ can induce transitions between the spin sub-levels \cite{Cohen}, or in other words, drive the Rabi oscillations. In the following, we will consider only that component of $\vec{B}_{MW}(t)$, and the directions of the three magnetic fields involved in the NV reference frame are shown in Fig. \ref{FigureS2}. We also want to stress that $\vec{B}_{MW}(t)$ is on resonance, meaning that $\omega = \omega_0$, where $\omega_0$ is the the frequency the $\ket{0} \leftrightarrow \ket{-1}$ transition $\omega_0 = D_{GS} - \gamma_{NV}B_0$. The time-dependent term of Eq. \ref{Eq_hamiltonian}, i.e. the interaction between the NVs and the oscillating MW field, can be simplified by switching to a reference frame $\left \{x_{R},y_{R},z_{NV}\right \}$ rotating with angular frequency $\omega$ around $z_{NV}$, where we assume that the $x_{R}$ axis is parallel to $\vec{B}_{LOOP}$. In this reference frame, both $\vec{B}_{LOOP}$ and $\vec{B}_{DUT}$ are static and, as they are on resonance with the $\ket{0} \leftrightarrow \ket{-1}$ transition, will lead to an inversion of the population between the spin sub-levels at a certain frequency $\Omega$. We first analyze the first measurement, ``MW loop'' of Fig. 1(f), where the only oscillating field is $\vec{B}_{LOOP}(t)$. In this case, the time dependent term $H_{MW}$ of the Hamiltonian of Eq. \ref{Eq_hamiltonian} in the rotating frame becomes:
\begin{equation}
  H_{MW} / h = \gamma_{NV} B_{LOOP} S_{x}
\label{Eq_hamiltonian_AC}
\end{equation}
and the Rabi oscillation between the $\ket{0} \leftrightarrow \ket{-1}$ states has a frequency\cite{Cohen} of:
\begin{equation}
  \Omega_{LOOP} = \gamma_{NV} |B_{LOOP}|
\label{Omega_loop}
\end{equation} 
When both the MW sources are driven (``MW loop + DUT'' measure of Fig. 1(f)), we need to take into account that there is a phase difference $\alpha$ between the two fields. In the rotating reference frame the two fields have the following form 
\begin{equation}
    \vec{B}_{LOOP} = (B_{LOOP},0,0) \hspace{10mm} \vec{B}_{DUT} = (B_{DUT} \cos \alpha,B_{DUT} \sin \alpha,0)
\label{B_L_and_B_D}
\end{equation}
and the term $H_{MW}$ of the Hamiltonian of Eq. \ref{Eq_hamiltonian} becomes:
\begin{equation}
  H_{MW} / h = \gamma_{NV} \left[ (B_{LOOP} + B_{DUT} \cos \alpha ) S_{x} +  (B_{DUT} \sin \alpha ) S_{y} \right ]
\label{Eq_hamiltonian_AC1}
\end{equation}
In this case the Rabi oscillation frequency is given by the vector sum of the two fields:
\begin{equation}
  \Omega_{LOOP+DUT} = \gamma_{NV} |\vec{B}_{MW}|  = \gamma_{NV} |\vec{B}_{LOOP}+\vec{B}_{DUT}|
\label{Omega}
\end{equation}
\begin{figure*}[b]
\centering
\includegraphics[width=0.95\linewidth]{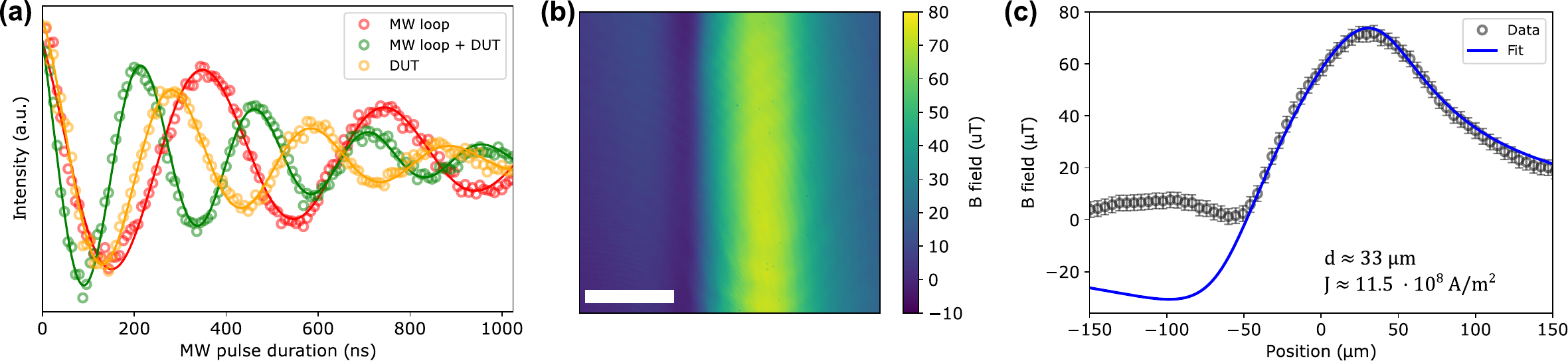}
\captionsetup{width=1\linewidth}
\caption{Results for wide-field imaging for an input power of the MW current in the DUT of 37.8 dBm. (a) Typical single-pixel Rabi oscillations data for “MW-loop” (red), “MW-loop + DUT” (green), and “DUT” (yellow). (b) Wide-field imaging of $B_{DUT}$ obtained with the differential measurement Eq. \ref{Eq_paper}. Scale bar is 100 $\upmu$m. (c) Magnetic field along an horizontal line-cut of (b) fit with Eq. \ref{B_fit_equation}.}
\label{FigureS3}
\end{figure*}
However, the two Rabi frequencies obtained from Eq. \ref{Omega_loop} and Eq. \ref{Omega} cannot be directly compared, as the Rabi oscillations occurs around two different axis, namely the one parallel to $\vec{B}_{LOOP}$ in the first case and the one parallel to $\vec{B}_{MW} = \vec{B}_{LOOP} + \vec{B}_{DUT}$ in the second. To be able to obtain information on $B_{DUT}$ from a differential measurement we need to first take into account that in our experimental conditions $|\vec{B}_{DUT}|\ll|\vec{B}_{LOOP}|$. This condition is fulfilled as in the ``DUT'' only measurement, no Rabi oscillations can be observed. With this condition we can just consider the projection of $\vec{B}_{DUT}$ on $\vec{B}_{LOOP}$ and the total field takes the following form $\vec{B}_{MW} \approx (B_{LOOP} + B_{DUT} \cos \alpha,0,0)$. The Rabi oscillation frequency in turn is:
\begin{equation}
  \Omega_{LOOP+DUT} \approx \gamma_{NV} |\vec{B}_{MW}|  = \gamma_{NV} |B_{LOOP}+B_{DUT} \cos \alpha|
\label{Omega_app}
\end{equation}
So when we perform the differential measurement between the ``MW loop + DUT'' and the ``DUT'' cases to isolate $B_{DUT}$, namely we take the difference between Eq. \ref{Omega_app} and Eq. \ref{Omega_loop}, we end up with:
\begin{equation}
  \Omega_{DUT} =  \Omega_{LOOP+DUT} -  \Omega_{LOOP} \approx \gamma_{NV} |B_{DUT} \cos \alpha|
\label{Omega_dut}
\end{equation}
Now we have to take into account that the phase difference $\alpha$ is random and can take any value in the interval $[-\pi,\pi]$ every time the measurement pulse sequence is run. For this reason, the signal we are actually measuring is the root mean square (RMS) of the above equation, resulting in:
\begin{equation}
\begin{aligned}
   \Omega_{DUT}^{RMS}  & = \sqrt{ \frac{1}{2\pi} \int_{-\pi}^{-\pi} ( \gamma_{NV} B_{DUT} \cos \alpha ) ^2 d\alpha} = \gamma_{NV} \sqrt{ \frac{1}{2\pi} \int_{-\pi}^{-\pi} B_{DUT}^2 \cos^2 \alpha d\alpha} \\
  & = \gamma_{NV} B_{DUT} \sqrt{ \frac{1}{2\pi} \int_{-\pi}^{-\pi} \frac{1-\cos\alpha}{2} d\alpha} = \frac{\gamma_{NV} B_{DUT}}{\sqrt{2}}
\label{B_eff_rms}
\end{aligned}
\end{equation}
In conclusion, putting together Eq. \ref{Omega_loop}, \ref{Omega_app}, \ref{Omega_dut}, and \ref{B_eff_rms}, we obtain:
\begin{equation}
  B_{DUT} = \sqrt{2}\left ( B_{LOOP+DUT} -B_{LOOP} \right )
\label{Eq_paper}
\end{equation}
that is Eq. 2 used in the main text.

\subsection{Particular case: $\mathbf{|B_{DUT}|  \lesssim |B_{LOOP}|}$}
The aim of this section it to prove that the differential measurement protocol used in this work are only works if the condition $|B_{DUT}| \ll |B_{LOOP}|$ is fulfilled. To do that, we analyze a wide-field Rabi oscillation measurement for an input MW power in the DUT of 37.8 dBm. We can observe that in this case the amplitude of the MW field generated by the DUT is strong enough to drive Rabi oscillations, as shown in the single-pixel measurement reported in \ref{FigureS3}(a). This leads to the conclusion that the amplitude $|B_{DUT}|$ is comparable to $|B_{LOOP}|$. The result obtained on $B_{DUT}$ through the differential measurement (Eq. \ref{Eq_paper}) are reported on Fig. \ref{FigureS3}(b) and Fig. \ref{FigureS3}(c) for respectively, the wide-field Rabi imaging and the fit of the data along an horizontal line-cut with Eq. \ref{B_fit_equation}. It can be easily observed that the model does not fit correctly the experimental data. This analysis proves that the condition $|B_{DUT}| \ll |B_{LOOP}|$ is necessary to get correct physical data from the differential measurement.

\section*{References}
\bibliography{Supplemental}